\documentclass[12pt]{article}
\usepackage{graphicx}
\usepackage{amssymb,amsmath, amsfonts,amsthm}
\usepackage{relsize}
\def\proof{\noindent\hspace{2em}{\itshape Proof: }}
\def\QEDclosed{\mbox{\rule[0pt]{1.3ex}{1.3ex}}} 

\def\QED{\QEDclosed} 
\def\endproof{\hspace*{\fill}~\QED\par\endtrivlist\unskip}

\input xy 
\xyoption{all} 

\usepackage{color}
\newtheorem{theorem}{Theorem}     
\newtheorem{definition}[theorem]{Definition}
\newtheorem{proposition}[theorem]{Proposition}
\newtheorem{lemma}[theorem]{Lemma}
\newtheorem{corollary}[theorem]{Corollary}

\newcommand{\eqa}{\begin{eqnarray}}
\newcommand{\eeqa}{\end{eqnarray}}
\newcommand{\beq}{\begin{equation}}
\newcommand{\eeq}{\end{equation}}
\newcommand{\nn}{\nonumber}

\def\d{\partial}
\def\n{\noindent}
\def\f{\frac}

\makeatletter
\@addtoreset{equation}{section}

\makeatother

\begin{document}
\title{Inherited structures in deformations of\\ Poisson pencils}
\author{Alessandro Arsie* and Paolo Lorenzoni**\\
\\
{\small *Department of Mathematics and Statistics}\\
{\small University of Toledo,}
{\small 2801 W. Bancroft St., 43606 Toledo, OH, USA}\\
{\small **Dipartimento di Matematica e Applicazioni}\\
{\small Universit\`a di Milano-Bicocca,}
{\small Via Roberto Cozzi 53, I-20125 Milano, Italy}\\
{\small *alessandro.arsie@utoledo.edu, **paolo.lorenzoni@unimib.it}}

\date{}

\maketitle

{\bf Abstract:}  In this paper we study some properties of bi-Hamiltonian deformations of Poisson pencils of hydrodynamic type.
 More specifically, we are interested in determining those structures of the fully deformed pencils that are inherited through
 the interaction between structural properties of the dispersionless pencils (in particular exactness or homogeneity) and suitable
 finiteness conditions on the central invariants (like polynomiality). This approach enables us to gain some information about each
 term of the deformation  {\em to all orders in} $\epsilon$. 

Concretely, we show that deformations of exact Poisson pencils of hydrodynamic type with polynomial central invariants can be put,
 via a Miura transformation, in a special form, that provides us with a tool to map a fully deformed Poisson pencil with polynomial
 central invariants of a given degree to a fully deformed Poisson pencil with constant central invariants to all orders in $\epsilon$.
 In particular, this construction is applied to the so called $r$-KdV-CH hierarchy that encompasses all known examples with non-constant
 central invariants. 

As far as homogeneous Poisson pencils of hydrodynamic type is concerned, we prove that they can also be put in a special form,
 if the central invariants are homogeneous polynomials. Through this we can compute the homogeneity degree about the tensorial
 component appearing in each order in $\epsilon$,  namely the coefficient of the highest order derivative of the $\delta$.


\section{Introduction}\label{sect0}
Toward the end of the twentieth century, Dubrovin and Zhang constructed a general framework
 to tackle the classification problem for integrable PDEs, motivated in part by questions arising
 in the theory of Gromov-Witten invariants and topological field theory. One of the cornerstones of their approach
 is the analysis and classification of Poisson pencils of the form
\begin{eqnarray}
\nonumber
&&\Pi^{ij}_{\lambda}=\omega^{ij}_{2}+\sum_{k\ge 1}\epsilon^k\sum_{l=0}^{k+1}A^{ij}_{(2)k,l}(q,q_x,\dots,q_{(l)})\delta^{(k-l+1)}(x-y)\\
\label{PP}
&&-\lambda\left(\omega^{ij}_{1}+\sum_{k\ge 1}\epsilon^k\sum_{l=0}^{k+1}A^{ij}_{(1)k,l}(q,q_x,\dots,q_{(l)})\delta^{(k-l+1)}(x-y)\right)
\end{eqnarray}
obtained via a bi-Hamiltonian deformation procedure from the dispersionless limit
$$\omega^{ij}_{2}-\lambda\omega^{ij}_{1}=g_{(2)}^{ij} \delta'(x-y)+\Gamma^{ij}_{(2)k} q^k_{x} \delta(x-y)-
\lambda\left(g_{(1)}^{ij} \delta'(x-y)+\Gamma^{ij}_{(1)k} q^k_{x} \delta(x-y)\right),$$
a so called Poisson pencil of hydrodynamic type  (here and in the subsequent Sections Einstein's convention about summed indices is enforced, unless stated otherwise). 
In their classification scheme, PDEs related via Miura transformations are considered equivalent, namely two pencils of the form \eqref{PP}
 are declared equivalent if they are obtained one from the other via a Miura transformation:
 \beq\label{Miura}
\tilde{q}^i= F_0^i(q)+\sum_{k\ge1}\epsilon^k F_k(q,q_x,\dots,q_{(k)}),\qquad{\rm det}\f{\d F_0^i}{\d q^j}\ne 0,\, {\rm deg}F^i_k=k.
\eeq
where, by definition ${\rm deg}\left(f(q)\right)=0$ and ${\rm deg}(q_{(k)})=k$. 
According to the theory, as further developed by Dubrovin, Liu and Zhang, once two compatible Poisson structures of hydrodynamic type $\omega_1$ and $\omega_2$ are chosen,
equivalence classes of Poisson pencils are labelled by
  $n$ functions called \emph{central invariants}. This set of invariants arise in the study
 of certain cohomology groups, called \emph{bi-Hamiltonian
cohomology groups} associated to the Poisson bivectors $\omega_1$ and $\omega_2$. 
 In order to define these cohomology groups one has to consider a double differential complex on the Grasmann algebra
 of multivector fields on the formal loop space $\mathcal{L}(\mathbb{R}^n)$. The two differentials of the complex, denoted by $d_{\omega_1}$ and $d_{\omega_2}$,
 are defined by
\begin{eqnarray}\label{diff1}
d_{\omega_1}&:=&[\omega_1,\cdot]:\Lambda^k\to\Lambda^{k+1}\\
\label{diff2}
d_{\omega_2}&:=&[\omega_2,\cdot]:\Lambda^k\to\Lambda^{k+1},
\end{eqnarray}
where the square bracket is the Schouten bracket and $\Lambda^k$ is the space of $k$-vector fields.
The Jacobi conditions
$$[\omega_1,\omega_1]=0,\qquad[\omega_2,\omega_2]=0$$
in this framework read
$$d_{\omega_1}^2=0,\qquad d_{\omega_2}^2=0.$$
Moreover due to the compatibility of $\omega_1$ and $\omega_2$:
$$[\omega_1,\omega_2]=0,$$
that is the differentials $d_{\omega}$ and $d_{\omega_2}$ anticommute. For much more information about these constructions see \cite{DZ}.

Since the deformed pencil \eqref{PP} is obtained from the pencil of hydrodynamic type via a complicated recursive procedure, in general it is very difficult to obtain 
information about the various orders of the deformation, and it is even more challenging to get insight about
 the complete deformed pencil at all orders in $\epsilon$. For instance,
 even the very existence of the deformation to all orders in $\epsilon$ has not been completely established in all cases
 (the existence of the deformed hierarchy and of one of the correspoding Poisson brackets has been solved with some additional assumptions
 coming from Gromov-Witten theory in \cite{BPS,BPS2}). 

The main contribution of our paper to this area is to look for those properties of the complete deformed pencil that are inherited through the interplay between special structural properties of the dispersionless limit (like being exact or homogeneous) and conditions on the central invariants (for instance a finiteness condition like being polynomial). This will enable us to prove some results for the deformed pencil {\em to all orders in} $\epsilon$ and to get some specific pieces of information about the tensorial component appearing in each order in $\epsilon$ (this component is just $A^{ij}_{(2)k,0}(q,q_x,\dots,q_{(l)})$, namely the coefficient of the highest order derivative of the $\delta$).

The key observation to implement this idea is to notice that, in general, those structural properties of the pencil $\omega_{\lambda}$ of hydrodynamic type that have a counterpart at the level of the bi-differential complex, combined with suitable conditions on the central invariants, give rise to special structures in the deformed pencil, that are natural to call {\em inherited structures}. It is not possible to expect that these structures depend only on the dispersionless pencil, because the isomorphism class (with respect to Miura transformations) of the fully deformed pencil depends on the choice of central invariants. For instance, in the situation in which the pencil $\omega_{\lambda}$ is exact, the choice of {\em constant} central invariants make the fully deformed pencil {\em exact} (see \cite{FL} and Theorem \ref{fund} for a generalization). Without this suitable assumption on the central invariants, exactness does not carry over to the fully deformed pencil.

In the present paper we will focus our attention to two cases, that appear to be particularly relevant due to their appearance in the framework of Frobenius manifolds and we will be mainly concerned with {\em polynomial} central invariants. 

The two cases are as follows:
\begin{enumerate}
 \item The Poisson pencil $\omega_{\lambda}=\omega_2-\lambda\omega_1$ is \emph{exact}.
 In this case there exists a vector field $e$ (sometimes called Liouville vector field)
 such that
\begin{eqnarray}
\label{exact1PP}
{\rm Lie}_e \omega_1&=&0\\
\label{exact2PP}
{\rm Lie}_e \omega_2&=&\omega_1.
\end{eqnarray}
At the level of the double differential complex the above properties imply
\begin{eqnarray}\label{qwer3}
\label{e1}
{\rm Lie}_e\,d_{\omega_1}-d_{\omega_1}{\rm Lie}_e&=&0\\
\label{e2}
{\rm Lie}_e\,d_{\omega_2}-d_{\omega_2}{\rm Lie}_e&=&d_{\omega_1}. 
\end{eqnarray} 
Combining \eqref{e1} and \eqref{e2} one obtains immediately
\beq\label{lieed1d2}
{\rm Lie}_e d_{\omega_1}\,d_{\omega_2}-d_{\omega_1}\,d_{\omega_2}{\rm Lie}_e=0. 
\eeq
\item The Poisson pencil $\omega_{\lambda}=\omega_2-\lambda\omega_1$ is \emph{homogeneous}. In this case there exists a vector field $E$
 such that
\begin{eqnarray}
\label{Hom1PP}
{\rm Lie}_E \omega_1&=&(d-2)\omega_1\\
\label{Hom2PP}
{\rm Lie}_E \omega_2&=&(d-1)\omega_2.
\end{eqnarray}
The properties  \eqref{Hom1PP} and \eqref{Hom2PP} imply
\begin{eqnarray}
\label{E1}
{\rm Lie}_E d_{\omega_1}-d_{\omega_1}{\rm Lie}_E&=&
(d-2)d_{\omega_1}\\
\label{E2}
{\rm Lie}_E d_{\omega_2}-d_{\omega_2}{\rm Lie}_E&=&
(d-1)d_{\omega_2}.
\end{eqnarray}
Combining \eqref{E1} and \eqref{E2} one gets
\beq\label{LieEd1d2}
{\rm Lie}_E d_{\omega_1}\,d_{\omega_2}-d_{\omega_1}\,d_{\omega_2}{\rm Lie}_E=(2d-3)d_{\omega_1}\,d_{\omega_2}.
\eeq
\end{enumerate}
As it was hinted above, 
part of the importance of these two cases stems from the fact that both instances arise in the theory
 of Frobenius manifolds. In this framework $e$ is the \emph{unity} vector field, $E$ is the \emph{Euler} vector field
 and $d$ is the \emph{charge} of the Frobenius manifold. More about this case will be presented in Section \ref{sect2} and \ref{sechomogeneous}. 
 
The paper is organized as follows. In Section \ref{sect2} we recall the main properties of the pencils of hydrodynamic type we will consider, namely those that are semisimple and bi-Hamiltonian. In particular, we present how the two structural properties of exactness and homogeneity appear at the level of the bi-differential complex and their relation with the theory of Frobenius manifolds. 

In Section \ref{sect3} the definition of central invariants is recalled and worked out in some examples. In particular, we recall the case of the $r$-KdV-CH hierarchy since it includes all known cases with non constant central invariants, and we are going to apply some of our results to it. In Section \ref{sect4} we briefly recall the notion of bi-Hamiltonian cohomology group, their significance in controlling the deformations of Poisson pencils and the fact that for a pencil like \eqref{PP} the deformation of $\omega_1$ can always be eliminated. 

In Section \ref{sectionexact}, we study those properties of the deformed pencil that are inherited from the exactness of the dispersionless limit, we work out some examples and we show how to apply our results to the case of the $r$-KdV-CH hierarchy with polynomial central invariants. We also show that deformations of exact Poisson pencils of hydrodynamic type with polynomial central invariants can be put, via a Miura transformation, in a special form, that we call normal form. In particular, this provides us with a tool to map a Poisson pencil with polynomial central invariants of a given degree to a Poisson pencil with constant central invariants to {\em all orders in }$\epsilon$. 

In Section \ref{sechomogeneous} we study deformations of homogeneous Poisson pencils of hydrodynamic type. We call homogenous a deformation $P_{\lambda}$ of a homogenous Poisson pencil of hydrodynamic type if its central invariants are homogenous functions of the \emph{same} degree $D$ in the canonical coordinates. In particular we prove that such a homogenous Poisson pencil $P_\lambda$ can be always reduced by a Miura transformation to a Poisson pencil $Q_\lambda$ of the form
$$
Q_{\lambda}=\omega_2+\sum_{k=1}^{\infty}
\epsilon^{2k}Q^{(2k)}_2-\lambda\omega_1
$$
such that $$
{\rm Lie}_E Q_{2}^{(2k)}=[(k+1)(d-1)+kD]Q_{2}^{(2k)},\qquad k=1,2,\dots$$
where $E$ is the Euler vector field. In particular, this allows us to predict exactly the homogeneity degree of the tensorial component in each term in $\epsilon$, prediction that we test on a nontrivial example. 

The short Section \ref{conclusions} provides some perspectives for future work. 



\section{The dispersionless case}\label{sect2}
Consider a semisimple bi-Hamiltonian structure of hydrodynamic type $\omega^{ij}_{\lambda}:=\omega^{ij}_2-\lambda\omega^{ij}_1$, where
$$\omega^{ij}_A=g^{ij}(q)_A\delta'(x-y)+\Gamma^{ij}_{A, k}(q)q^k_x\delta(x-y), \quad A=1,2.$$
This means that
\begin{enumerate}
 \item Each $\omega^{ij}_A$ defines a Poisson bivector, which on the other hand, under the assumption ${\rm det}(g^{ij}_A)\neq 0$ is equivalent to $g^{ij}_A$ being flat and $\Gamma^{ij}_{A, k}=-g^{il}_A \Gamma^{j}_{A, lk}$, where $ \Gamma^{j}_{A, lk}$ are the Christoffel symbols of $g_{A,ij}$
 (see \cite{DN}).
 \item The Riemann tensor of the pencil $g^{ij}_{\lambda}=g^{ij}_2-\lambda g_1^{ij}$
 vanishes for any value of $\lambda$ and the Christoffel symbols $\Gamma^{ij}_{\lambda,k}$ of the pencil
 $g^{ij}_{\lambda}$ are $\Gamma^{ij}_{(2),k}-\lambda \Gamma^{ij}_{(1),k}$
 (this is equivalent to the two Poisson structures $\omega_1$ and $\omega_2$ being compatible
 and thus defining a bi-Hamiltonian structure, see \cite{D07})
 \item The roots $u^1(q), \dots, u^n(q)$ of the characteristic equation ${\rm det}g_{\lambda}={\rm det}(g_2-\lambda g_1)=0$ are functionally independent (this is the condition of $g_{\lambda}$ being semisimple). 
 \end{enumerate}
Due to the fact that the bi-Hamiltonian structures we are dealing with are semisimple, we can re-express the quantities involved in term of the functions $u^i(q)$ which are called {\em canonical coordinates}.  It can be proved that both metrics $g_1$ and $g_2$ are in diagonal form using canonical coordinates \cite{Fe}: 
 \begin{equation}\label{eqmetric}g_1^{ij}=f^i(u)\delta_{ij}, \quad g_2^{ij}=u^i f^i(u) \delta_{ij}.\end{equation}

A special class of Poisson pencils $\omega_{\lambda}$ is given by \emph{exact Poisson pencil}. In this case we have the following theorems
\begin{theorem}\cite{FL}
A semisimple bi-Hamiltonian structure of hydrodynamic type is exact if and only if the functions $f^i(u)$ in \eqref{eqmetric}
satisfy the condition 
\begin{equation}\label{essential}
\sum_{k=1}^n \frac{\partial f^i(u)}{\partial u^k}=0, \quad i=1, \dots, n\end{equation}
Moreover, in canonical coordinates all the components of the vector field $e$ are equal to $1$. 
\end{theorem}

How the fact that the Poisson pencil is exact translates at the level of cohomology operators is spelled out in the following:
\begin{theorem}\label{commutativity}
Let $d_{\omega_1}$ and $d_{\omega_2}$ be the Poisson cohomology differentials associated to a semisiple exact
 Poisson pencil $\omega_{\lambda}$. Then the relations \eqref{e1}
 and \eqref{e2} are satisfied.
\end{theorem}
\proof
Let $\Lambda$ be a $k$-multivector. Then $({\rm Lie}_e\circ d_{\omega_1})(\Lambda)=[e, [\omega_1, \Lambda]]$, where $[\cdot, \cdot]$ denotes the Schouten-Nijenhuis bracket. By the graded Jacobi identity, the following identity holds: 
$$(-1)^{k-1}[e, [\omega_1, \Lambda]]+[\omega_1, [\Lambda, e]]+(-1)^k[\Lambda, [e, \omega_1]]=0.$$
Since ${\rm Lie}_e\omega_1=0$, and $[\Lambda, e]=(-1)^k [e, \Lambda]$, the graded Jacobi identity reduces to $(-1)^{k-1}[e, [\omega_1, \Lambda]]+(-1)^k[\omega_1, [e,\Lambda]]=0$, which is exactly \eqref{e1}. For \eqref{e2} the proof is entirely analogous, just observe that in this case the term $[e, \omega_2]=\omega_1$ and $[\Lambda, \omega_1]=(-1)^{2k}[\omega_1, \Lambda]$. 
\endproof

Relation \eqref{e1} means that ${\rm Lie}_e$ is a $d_{\omega_1}$-chain map, 
so it descends to a map in the $d_{\omega_1}$ Poisson cohomology,
 while equation \eqref{e2} intertwines the action of $d_{\omega_1}$ and $d_{\omega_2}$ via ${\rm Lie}_e$.
 In particular, it is immediate to see that \eqref{e2} implies that ${\rm Lie}_e$ descends to a map at the level of the bi-Hamiltonian cohomology groups. It is also true that ${\rm Lie}_e$ induces maps $H^*(d_{\omega_2})$ to $H^*(d_{\omega_1})$, however as such this is not interesting, since it is already known that $H^*(d_{\omega_i})=0$ $i=1,2$.

An other case we are going to deal with is a special case of homogeneous Poisson pencils
 defined by \eqref{Hom1PP} and \eqref{Hom2PP}. We will additionally assume that the pencil $g_{\lambda}$
 is semisimple and that, in canonical coordinates, the Euler vector field $E$ is given by the formula
\beq\label{EulerVF}
E=\sum_{i=1}^n u^i\f{\d}{\d u^i}.
\eeq
The motivation for this assumption comes from the theory of Frobenius manifolds. In such a context
 we have two flat metrics $g_1$ and $g_2$ satisfying
 the properties 
\begin{eqnarray}
\label{Hom1}
{\rm Lie}_E g_1^{ij}&=&(d-2)g^{ij}_1\\
\label{Hom2}
{\rm Lie}_E g_2^{ij}&=&(d-1)g^{ij}_2.
\end{eqnarray}
Moreover the contravariant components of the second metric, the so-called \emph{intersection form}, are given by
\beq\label{intform}
g_2^{ij}=g_1^{il}c^j_{lk}E^k
\eeq
where $c^j_{lk}$ are the structure constants defining the Frobenius algebra on the tangent spaces. 
 The special form of the Euler vector field in canonical coordinates follows immediately from 
 the formula \eqref{intform} and the semisimplicity assumption. From \eqref{Hom1} and \eqref{Hom2} it follows that in canonical coordinates 
 the contravariant components of the metric $g_1$ are homogeneous
 functions of degree $d$ and the contravariant components of the metric $g_2$ are homogenous
 functions of degree $d+1$. We show now that the Poisson pencil associated to such metrics is homogeneous. Indeed:
\begin{eqnarray*}
&&{\rm Lie}_{E}\omega_1^{ij}=\\
&&\sum_{k,s}\left( \partial^s_x E^k(u(x), \dots) 
\frac{\partial \omega_1^{ij}}{\partial u^k_{(s)}(x)}
- \frac{\partial E^i(u(x), \dots)}{\partial u^k_{(s)}(x)} \partial ^s_x \omega_1^{kj}
-\frac{\partial E^j(u(y), \dots)}{\partial u^k_{(s)}(y)}\partial^s_y \omega_1^{ik}\right)=\\
&&\sum_k u^k 
\frac{\partial \omega_1^{ij}}{\partial u^k}+
\sum_k u^k_x 
\frac{\partial \omega_1^{ij}}{\partial u^k_x}
-  2\omega_1^{ij}=\\
&&(d-2)g_1^{ij}\,\delta'(x-y)
+\sum_k u^k 
\frac{\partial \Gamma_{(1)l}^{ij}}{\partial u^k}u^l_x\,\delta(x-y)
-\sum_k \Gamma_{(1)k}^{ij}u^k_x\,\delta(x-y)=\\
&&(d-2)\left[g_1^{ij}\,\delta'(x-y)+\sum_k \Gamma_{(1)k}^{ij}u^k_x\,\delta(x-y)\right]
=(d-2)\omega^{ij}
\end{eqnarray*}
where, in the last identity, we used the fact that the Christoffel symbols $\Gamma_{(1)k}^{ij}$ are homogeneous functions of degree $d-1$. Similarly
 using the  fact that the Christoffel symbols $\Gamma_{(2)k}^{ij}$ are homogeneous functions of degree $d$ one obtains \eqref{Hom2PP}.
 The identities \eqref{E1} and \eqref{E2} can be easily proved using graded Jacobi identity as in Theorem \ref{commutativity}.

\section{Central invariants}\label{sect3}
In the semisimple case \cite{LZ} (that is if $\omega_{\lambda}$ is semisimple) equivalence classes
 of  equivalent Poisson pencil of the form \eqref{PP} are labelled by $n$ functional
 parameters called \emph{central invariants}. More precisely two pencils having
 the same leading order are Miura equivalent if and only if they have
 the same central invariants. The problem of costructing a pencil for a given
 choice of the leading term $\omega_{\lambda}$ and of 
the central invariants has been solved only in certain cases. In general, as observed in the Introduction, even to prove the existence of the pencil is a non trivial problem. The central invariants are defined as
\beq\label{altdef}
c_i=-\f{1}{3f^i}{\rm Res}_{\lambda=u^i}{\rm Tr}\,g^{-1}_{\lambda}A_{\lambda}
\eeq
where the tensor $A^{ij}$ is defined by
$$ A^{ij}_{\lambda}=A^{ij}_{(\lambda)2,0}+(g_{\lambda}^{-1})_{lk}A_{(\lambda)1,0}^{li}
A_{(\lambda)1,0}^{kj}.
$$
with
$$A^{ij}_{(\lambda)2,0}=A^{ij}_{(2)2,0}-\lambda A^{ij}_{(1)2,0},\qquad A^{ij}_{(\lambda)1,0}=A^{ij}_{(2)1,0}-\lambda A^{ij}_{(1)1,0}.$$
It turns out \cite{LZ,DLZ} that the central invariants $c_i$ depend only on the canonical coordinates $u^i$  and are given by the following expression:
\beq\label{CInv}
c_i(u^i)=\f{1}{3(f^i)^2}\left(Q_2^{ii}-u^i\,Q_1^{ii}+\sum_{k\ne i}\f{(P_2^{ki}-u^i\,P_1^{ki})^2}{f^k(u^k-u^i)}\right),\,\,\,i=1,\dots,n.
\eeq
where $P_1^{ij},\,P_2^{ij},\,Q_1^{ij},\,Q^{ij}_2$ are the components of the tensor fields
$A^{(1)ij}_{2,0},\,A^{(2)ij}_{2,0}$, $A^{(1)ij}_{3,0},\,A^{(2)ij}_{3,0}$ in canonical coordinates.
 In particular, each central invariant $c_i$ is a scalar function of only the canonical coordinate $u^i$. 

Now we show how the definition of central invariants works in a couple of examples:
\paragraph {{AKNS.}} 
Let us consider the Poisson pencil $\omega_2+\epsilon P_2^{(1)}-\lambda \omega_1$ with
\beq\label{AKNS}
\omega_2+\epsilon P_2^{(1)}-\lambda \omega_1=
\begin{pmatrix}
          (2u\partial_x+u_x)\delta &  v\delta'\\ 
          \partial_x(v\delta) & -2\delta'
         \end{pmatrix}+\epsilon\begin{pmatrix}
          0 &  -\delta''\\ 
\delta'' & 0
         \end{pmatrix}
-\lambda\begin{pmatrix}
          0 & \delta'\\ \delta' & 0 
         \end{pmatrix}
\eeq
where, to keep formulas short, we write $\delta$ instead of $\delta(x-y)$ and $\delta'$ is derivative with respect to $x$. 
This is the Poisson pencil of the so-called AKNS (or two-boson) hierarchy.

In this case
$$g_{\lambda}=\begin{pmatrix} 2u&v-\lambda\\v-\lambda
&-2\end{pmatrix}.$$
After some computations we get $A_{\lambda}=\f{g_{\lambda}}{{\rm det}g_{\lambda}}$ 
and therefore, taking into account that
$$u^1=v+\sqrt{-4u},\, u^2=v-\sqrt{-4u}.\qquad f^1=\frac{8}{u_2-u_1},\, f^2=\frac{8}{u_1-u_2},$$
 using formula \eqref{altdef} we obtain 
\begin{eqnarray*}
 c_1&=&-\f{1}{3f^1}{\rm Res}_{\lambda=u^1}{\rm Tr}\,g^{-1}_{\lambda}A_{\lambda}=
-\f{1}{3f^1}{\rm Res}_{\lambda=u^1}\f{2}{{\rm det}g_{\lambda}}=-\f{1}{12}\\
c_2&=&-\f{1}{3f^2}{\rm Res}_{\lambda=u^2}{\rm Tr}\,g^{-1}_{\lambda}A_{\lambda}=
-\f{1}{3f^2}{\rm Res}_{\lambda=u^2}\,\f{2}{{\rm det}g_{\lambda}}=-\f{1}{12}
\end{eqnarray*}

\paragraph{{Two component CH.}} 
Moving $P_2^{(1)}$ from $P_2$ to $P_1$ in the Poisson pencil of the AKNS hierarchy one obtains the following
 Poisson pencil \cite{F,LZ}
\beq\label{PP2CH}
P_{\lambda}
=\begin{pmatrix}
          (2u\partial_x+u_x)\delta &  v\delta'\\ 
          \partial_x(v\delta) & -2\delta'
         \end{pmatrix}-\lambda\begin{pmatrix}
          0 & \delta' -\epsilon\delta''\\ \delta'+\epsilon\delta'' & 0 
         \end{pmatrix}
\eeq
which is the Poisson pencil that identifies the so called CH$_2$ hierarchy.
The pencil $g_{\lambda}$ and the canonical coordinates are the same as in the previous example, while
 $A_{\lambda}=\f{\lambda^2 g_{\lambda}}{{\rm det}g_{\lambda}}$.  Using formula \eqref{altdef} we obtain
\begin{eqnarray*}
 c_1&=&-\f{1}{3f^1}{\rm Res}_{\lambda=u^1}{\rm Tr}\,g^{-1}_{\lambda}A_{\lambda}=
-\f{1}{3f^1}{\rm Res}_{\lambda=u^1}\f{2\lambda^2}{{\rm det}g_{\lambda}}=-\f{(u^1)^2}{12}\\
c_2&=&-\f{1}{3f^2}{\rm Res}_{\lambda=u^2}{\rm Tr}\,g^{-1}_{\lambda}A_{\lambda}=
-\f{1}{3f^2}{\rm Res}_{\lambda=u^2}\,\f{2\lambda^2}{{\rm det}g_{\lambda}}=-\f{(u^2)^2}{12}.
\end{eqnarray*}

Both the above examples are special cases of the $r$-KdV-CH hierarchy.

\paragraph{{$r$-KdV-CH hierarchy.}}

The $r$-KdV-CH hierarchy is an encompassing generalization of the Kortweg-de-Vries and Camassa-Holm
hierarchies parameterized by $r+1$ constants. It has been introduced by Antonowicz and Fordy in \cite{AF87} and \cite{AF88} (see also
 \cite{MA}) and further studied by Chen, Liu and Zhang (\cite{CLZ09}). It appears that the only known bi-Hamiltonian hierarchies with non-constant central invariants are special cases of the $r$-KdV-CH hierarchy. For details about the $r$-KdV-CH hierarchy we refer to \cite{CLZ09}, here we just focus our attention on its bi-Hamiltonian structure (it is actually multi-Hamiltonian) and how our results can be applied in this case. 

Fix $r+1$ constants $(a_0, a_1, \dots, a_r)\in (\mathbb{R}^{r+1})^*$, so $(a_0, \dots, a_r)\neq (0,\dots, 0)$, and coordinates $w^0, \dots, w^{r-1}$ on a manifold $M$, which is considered the target manifold for the loop space. The  $r$-KdV-CH hierarchy is endowed with $r+1$ mutually compatible Hamiltonian structures 
\begin{equation}\label{hamiltonian1r}
(P_m)^{ij}=f^{ij}_m \mathcal{D}_{i+j+1-m}, \quad i,j=0, \dots, r-1, \quad m=0, \dots, r
\end{equation}
where $$\mathcal{D}_{i}=2w^i\partial_x +w^i_x-\frac{\epsilon^2}{2} a_i\partial^3_x, \quad w^r:=1, \quad w^i=a_i:=0, \quad \text{for } i<0,\; i>r$$
and where furthermore 
$$f^{kl}_m:=\left\{ 
\begin{array}{cc}
1 & l<m \text{ and } k<m\\
-1 & l\geq m \text{ and } k\geq m\\
0 & \text{in all other cases} 
\end{array}\right.$$
Among the mutually compatible Hamiltonian structures $P_m$, $m=0, \dots, r$, we focus our attention on two of them, say $P_k$ and $P_l$. 
Their dispersionless limit is given by 
$$Q^{ij}_b=P^{ij}_{b,{|\epsilon=0}}=f^{ij}_b(2w^{i+j+1-b}\partial_x+w^{i+j+1-b}_x), \quad b=k,l.$$
To study the dispersionless bi-Hamiltonian structure $(Q_k, Q_l)$ one introduces the following coordinates. Consider the polynomial $P(\lambda)=\lambda^r+w^{r-1}\lambda^{r-1}+\dots+w^0$. If its roots $\lambda_i$ are pairwise distinct, which is equivalent to $P'(\lambda_i)\neq0$, then they can be used as a local system of coordinates, in place of $w^i$. Let us remind that Chen, Liu and Zhang (see \cite{CLZ09}) proved that in this case the bi-Hamiltonian structure $(Q_k, Q_l)$ is semisimple and has canonical coordinates $u^i:=(\lambda_i)^{l-k}$, $i=1, \dots, r$ and that in these coordinates, the associated metrics $g_k$ $g_l$ have non-zero components given by $$g^{ii}_k=-2(k-l)^2\frac{{\lambda_i}^{2l-k-2}}{P'(\lambda_i)}, \quad g^{ii}_m=-2(k-l)^2\frac{{\lambda_i}^{3l-2k-2}}{P'(\lambda_i)}, \quad i=1, \dots, r.$$
Although the computation of the central invariants $c^i(u^i)$ for the pair $(P_k,P_l)$ has appeared in Chen,
 Liu and Zhang, for convenience of the reader and since this example will be worked out later we report here the details of such a computation. 
Given $P_b=Q_b(w)+\epsilon^2(E(w)_b\partial^3_x+\dots)$, $b=k,l$ as above, the $i$-th central invariant $c^i$ which is just a function of the canonical coordinate $u^i$ is given by \begin{equation}\label{hamiltonianr3}
c^i(u^i)=\frac{E^{ii}(u)_l-u^iE^{ii}(u)_k}{3(g^{ii}(u)_k)^2}.
\end{equation}
Equivalently, since $u^i=(\lambda_i)^{l-k}$, we can express \eqref{hamiltonianr3} as a function of $\lambda_i$. 
Now $E^{ij}_m(w)=-\frac{1}{2}f^{ij}_ma_{i+j+1-m}$ and therefore
$$E^{ii}_m(\lambda)=-\frac{1}{2}\sum_{k,l=0}^{r-1}f^{kl}_ma_{k+l+1-m}\frac{\partial \lambda_i}{\partial w^k}\frac{\partial \lambda_i}{\partial w^l},$$
and using the definition of $f^{ij}_m$ this is equal to 
\begin{equation}\label{hamiltonianr5}E^{ii}_m(\lambda)=-\frac{1}{2}\sum_{k,l=0}^{m-1}a_{k+l+1-m}\frac{\partial \lambda_i}{\partial w^k}\frac{\partial \lambda_i}{\partial w^l}+\frac{1}{2}\sum_{k,l=m}^{r-1}a_{k+l+1-m}\frac{\partial \lambda_i}{\partial w^k}\frac{\partial \lambda_i}{\partial w^l}.\end{equation}
Moreover, the following relation holds:\begin{equation}\label{hamiltonianr4}\frac{\partial \lambda_i}{\partial w^k}=-\frac{\lambda^k_i}{P'(\lambda_i)}.\end{equation}
Indeed, since $P(\lambda)=\lambda^r+w^{r-1}\lambda^{r-1}+\dots+w^0=\prod_{i=1}^r(\lambda-\lambda_i)$, we have
$$\frac{\partial P(\lambda)}{\partial w_k}=\lambda^k=-\sum_l\left(\prod_{j\neq l}(\lambda-\lambda^j)\frac{\partial \lambda_l}{\partial w_k}\right).$$
Evaluating this last expression at $\lambda=\lambda_i$ we get immediately 
$\lambda^k_i=-\prod_{j\neq i}(\lambda_i-\lambda_j)\frac{\partial \lambda_i}{\partial w_k}$ and so 
$$\frac{\partial \lambda_i}{\partial w_k}=-\frac{\lambda^k_i}{\prod_{j\neq i}(\lambda_i-\lambda_j)}.$$
Since $P'(\lambda)=\sum_l\left(\prod_{j\neq l}(\lambda-\lambda_j) \right)$, we arrive at equation \eqref{hamiltonianr4}.
Substituting \eqref{hamiltonianr4} into \eqref{hamiltonianr5} we arrive at the expression 
\begin{equation}\label{hamiltonianr6}
E^{ii}_m(\lambda)=-\frac{1}{2}\frac{1}{(P'(\lambda_i))^2}\left(\sum_{k,l=0}^{m-1}\lambda^{k+l}_i a_{k+l+1-m}-\sum_{k,l=m}^{r-1}\lambda^{k+l}_ia_{k+l+1-m} \right)
\end{equation}
Defining the polynomial $\frak{a}(\lambda):=a_0+a_1\lambda+\dots+a_r\lambda^r$, rearranging the sums in the right hand side of \eqref{hamiltonianr6}, it is not difficult to see that 
\begin{equation}\label{hamiltonianr7}
E^{ii}_m(\lambda)=\frac{\lambda^m_i \frak{a}'(\lambda_i)-m \lambda^{m-1}_i \frak{a}(\lambda_i)}{2(P'(\lambda_i))^2}.
\end{equation}
Now observe that $$E^{ii}_m(u)=\left(\frac{\partial u^i}{\partial \lambda_i} \right)^2 E^{ii}_m(\lambda),$$ so since $u^i=(\lambda_i)^{l-k}$ we get
\begin{equation}\label{hamiltonianr8}E^{ii}_m(\lambda_i(u^i))=(l-k)^2\lambda^{2(l-k-1)}\frac{\lambda^m_i \frak{a}'(\lambda_i)-m \lambda^{m-1}_i \frak{a}(\lambda_i)}{2(P'(\lambda_i))^2}.\end{equation}
To focus on the central invariants for the pair $(P_k, P_l)$ so let us assume $l>k$ and use formula \eqref{hamiltonianr3} to compute the central invariants; using \eqref{hamiltonianr8} and the fact that $u^i=(\lambda_i)^{l-k}$ we get:
$$c^i(u^i)=\frac{E^{ii}_l-\lambda^{l-k}_iE^{ii}_k}{3(g^{ii}_k)^2}=\frac{\lambda^{1-l}_i\frak{a}(\lambda_i)}{24(k-l)}=\frac{(u^i)^{\frac{1-l}{l-k}} \frak{a}((u^i)^{\frac{1}{l-k}}) }{24(k-l)}.$$
Let us remark that the central invariants $c^i(u^i)$ for the compatible pair $(P_k, P_l)$ of the $r$-Kdv-CH hierarchy are rational functions of the canonical coordinates if and only if $l=k+1$. It is immediate to check that the condition is sufficient. To show that it is necessary, observe that $c^i(u^i)$ are rational if and only if $\frac{1-l}{(l-k)^{r+1}}\in \mathbb{Z}$. If $l=1$ then the condition is satisfied and $k=0$. Otherwise, call $q=l-k$, $1\leq q \leq r$. Now if $q\geq 2$, then $q^{r+1}>|1-r|>|1-l|$ for $2\leq l \leq r$, and this shows that the ratio we are dealing with can be integer if and only if $q=1$, which means $l=k+1$. Moreover, $c^i(u^i)$ are polynomials in the canonical coordinates if and only if $k=0, l=1$. In the next few sections we will use the $r$-KdV-CH hierarchy as an example on which to illustrate some of our results.

\section{Bi-Hamiltonian cohomology}\label{sect4}
For the sake of being reasonably self-contained and to fix notations,
 in this section we collect some definitions and results about (bi)-Hamiltonian cohomologies
 and the Dubrovin-Zhang complex (see \cite{DZ} for full details and proofs).
 Let $g$ be a flat metric on $\mathbb{R}^n$ and
 $\omega$ be the associated Poisson bivector of hydrodynamic type. In analogy with
 the case of finite dimensional Poisson manifolds \cite{lichn} one defines 
 Poisson cohomology groups in the following way: 
\begin{equation}\label{eq33bis.eq}
H^j(\mathcal{L}(\mathbb{R}^n), \omega):=\frac{\ker\{d_{\omega}: \Lambda^j_{\text{loc}}\rightarrow \Lambda^{j+1}_{\text{loc}}\}}
{\mathrm{im}\{d_{\omega}: \Lambda^{j-1}_{\text{loc}} \rightarrow \Lambda^j_{\text{loc}}\}}
\end{equation}
where $d_{\omega}:=[\omega,\cdot]$ (the square brackets denote the Schouten brackets) 
and $\Lambda^{j}_{\text{loc}}$ is the space of local $j$-multivectors
on the loop space $\mathcal{L}(\mathbb{R}^n)$ (see \cite{DZ}
 for more details on the definition of this complex). 
 Since $\Lambda^{j}_{\text{loc}}$ has a natural decomposition in homogenous components  
 which is preserved by $d_{\omega}$,  we have
\begin{equation}\label{eq34.eq}
H^j(\mathcal{L}(\mathbb{R}^n), \omega)=\oplus_{k} H^j_k(\mathcal{L}(\mathbb{R}^n), \omega).
\end{equation}
For Poisson structures of hydrodynamic type, it has been proved in \cite{G} (see also \cite{DMS} for an independent proof
 of the cases $n=1,2$) that $H^k(\mathcal{L}(\mathbb{R}^n), \omega)=0$ for $k=1,2,\dots$. 
  The vanishing of these cohomology groups implies that any deformation of a Poisson bivector of hydrodynamic type  
\begin{equation}\label{DPB}
P^{\epsilon} =\omega +\sum_{n=1}^{\infty}\epsilon^{n}P_{n},
\end{equation}
where $P_k\in \Lambda^2_{k+2, \text{loc}}$ can be obtained from $\omega$ by performing a Miura transformation. 

In order to study deformations of Poisson pencils  of hydrodynamic type  it is necessary to introduce
 bi-Hamiltonian cohomology groups \cite{GZ,DZ,LZ} 
 They are defined as
$$H^2_{k}(\mathcal{L}(\mathbb{R}^n),\omega_1,\omega_2)
=\f{{\rm Ker}\left(d_{\omega_1} d_{\omega_2}\,|_{\Lambda^1_{k,\text{loc}}}\right)}{{\rm Im}
\left(d_{\omega_1}|_{\Lambda^0_{k-2,\text{loc}}}\right)\oplus{\rm Im}\left(d_{\omega_1}|_{\Lambda^0_{k-2,\text{loc}}}\right)}.$$  
Liu and Zhang showed that, in the semisimple case,
$$H^2_{k}(\mathcal{L}(\mathbb{R}^n),\omega_1,\omega_2)=0
\quad \forall k\ne 2,$$
and that the elements of
$$H^2_{2}(\mathcal{L}(\mathbb{R}^n),\omega_1,\omega_2)$$
have representatives of the form
\begin{equation}\label{qtvf}
d_2\left(\sum_{i=1}^n\int c^i(u^i)u^i_x{\rm log}u^i_x\,dx\right)-d_1\left(\sum_{i=1}^n\int u^i c^i(u^i)u^i_x{\rm log}u^i_x\,dx\right)
\end{equation}
where $c^i(u^i)$ are the central invariants introduced
 in the previous section.  More explicitly, the components of these vector fields,
 in canonical coordinates, are given by
\beq\label{LZVF}
X^i=\sum_{j=1}^n\left[\left(\frac{1}{2}
\delta_{ij}\partial_x f^i + A^{ij}\right)
c^j u^j_x + (2\delta_{ij}f^i - L^{ij})\partial_x(
c^j u^j_x)\right],\,i=1,\dots,n.
\eeq
with
\begin{eqnarray} 
A^{ij}&=&\frac{1}{2}\label{A}
\left(\frac{f^i}{f^j}\frac{\d f^j}{\d u^i} u^j_x -\frac{f^j}{f^i} \frac{\d f^i}{\d u^j} u^i_x\right)\\
\label{L}
L^{ij}&=&\frac{1}{2}\delta_{ij}f^i +\frac{(u^i - u^j)f^i}{2f^j}
\frac{\d f^j}{\d u^i}.
\end{eqnarray}

We will use these facts later.

\section{Deformations of exact Poisson pencil
 of hydrodynamic type}\label{sectionexact}
In this section we study deformations of exact
 Poisson pencil of hydrodynamic type. In the case 
 of costant central invariants we have the following theorem \cite{FL}.

\begin{theorem}\label{FL2}
Suppose we are given a Poisson pencil 
\begin{equation}\label{poissontot}
\Pi_{\lambda}=P_2-\lambda P_1=\omega_2+\sum_{k=1}^{\infty} \epsilon^{2k} P^{(2k)}_2-\lambda\left(\omega_1+\sum_{k=1}^{\infty} \epsilon^{2k} P^{(2k)}_1 \right),
\end{equation}
whose $\epsilon=0$ limit $\omega_{\lambda}=\omega_2-\lambda \omega_1$ is an exact semisimple bi-Hamiltonian structure of hydrodynamic type.
 Then the central invariants of $\Pi_{\lambda}$ are constant functions of the canonical coordinates if and only if $\Pi_{\lambda}$ is an exact Poisson pencil.  
\end{theorem}

In particular in the proof of Theorem \ref{FL2}, it is shown that there exists a Miura transformation reducing the Poisson pencil $\Pi_{\lambda}$ \eqref{poissontot} to the form \begin{equation}\label{poisson}
\Pi_{\lambda}=\omega_{\lambda}+\sum_{k=1}^{\infty} \epsilon^{2k}P^{(2k)}_2=\omega_2+\sum_{k=1}^{\infty} \epsilon^{2k}P^{(2k)}_2-\lambda \omega_1
\end{equation}
and there exists a Miura transformation preserving $\omega_1$ and reducing the vector field $Z$ involved
 in the exactness of $\Pi_{\lambda}$ to $e=\sum_{k=1}^n \frac{\partial}{\partial u^k}$.
 In this system of coordinates, the exactness of the pencil $\omega_{\lambda}$ is expressed
 as ${\rm Lie}_eP_2=\omega_1$ and ${\rm Lie}_e\omega_1=0$, where $P_2:= \omega_2+\sum_{k=1}^{\infty} \epsilon^{2k}P^{(2k)}_2.$

Let us also recall that all the terms $P^{(2k)}_2$ in \eqref{poisson} are co-boundaries of $\omega_1$,
 so they can be written as Lie derivatives of $\omega_1$ with respect to certain vector fields.
 In particular, the first term in the deformation of $\Pi_{\lambda}$, namely $P^{(2)}_2$ can always be chosen to be  
\begin{equation}\label{p22}
P^{(2)}_2={\rm Lie}_{X}\omega_1
\end{equation}
where the vector field $X$ has components $X^i$ in canonical coordinates given by \eqref{LZVF}. We say that the first order term in $\epsilon$ in the deformation of $\Pi_{\lambda}$, namely $P^{(2)}_2$ is in {\em standard form} if it is chosen as \eqref{p22}.

Moreover, the vector field $X$ explicitly written as a vector field on the loop manifold $\mathcal{L}(\mathbb{R}^n)$ is given by 
$$X=\sum_{i=1}^n \sum_{s\geq 0}\partial^s_x X^i \frac{\partial }{\partial u^i_{(s)}}.$$

At this point we can prove the following, which generalizes the conditions about the constancy of the central invariants proved in \cite{FL}:

\begin{proposition}\label{centralinvariantsn}
Let $\Pi_{\lambda}$ be a bi-Hamiltonian structure as in \eqref{poisson} whose dispersionless 
limit $\omega_{\lambda}$ is an exact semisimple bi-Hamiltonian structure of hydrodynamic type
 and with $P^{(2)}_2$ in standard form.  
Then 
\beq\label{maxdeg}
{\rm max}_{i=1,\dots,n}{\rm deg}\,c_i(u)=l
\eeq
if and only if the following conditions are satisfied: 
\begin{equation}\label{homogeneity}
{\rm Lie}^{l+1}_eP^{(2)}_2=0, \quad {\rm Lie}^{l}_e P^{(2)}_2\neq0.
\end{equation}
\end{proposition}
\proof
First we prove that \eqref{maxdeg} implies  \eqref{homogeneity}. 
Using the formulas provided in [DZ] to compute ${\rm Lie}_e X_{c_1,\dots,c_n}$, 
where $X_{c_1,\dots,c_n}$ is given in \eqref{LZVF}, we find 
$$\left({\rm Lie}_e X_{c_1,\dots,c_n}\right)^i=[e, X_{c_1,\dots,c_n}]^i
=\sum_{h=1}^n \frac{\partial X^i_{c_1,\dots,c_n}}{\partial u^h}.$$
Call $D_e$ the differential operator $\sum_{h=1}^n  \frac{\partial \cdot}{\partial u^h}$
 acting on the components $X^i$ given by the formula \eqref{LZVF}.
 Since by the exactness of the pencil $\omega_{\lambda}$ we have that $D_e(f^i)=0$ (\eqref{essential}),
 it is immediate to check that similarly $D_e(A^{ij})=D_e(L^{ij})=0$.
 Moreover, $D_e$ and $\partial_x$ commute, 
therefore when $D_e$ acts on $X^i$ is just acting on the central invariants $c^i(u)$. 
 This reasoning can be repeated for any iteration of the Lie derivative ${\rm Lie}_e$;
 in particular the $i$-th component of the $k$-th iterated  Lie derivative $[{\rm Lie}^k_e(X)]^i$ is just given by the action of the differential operator $D^k_e$  on $X^i$. So using the notation just introduced, we have that 
$$[{\rm Lie}^k_e X_{c_1,\dots,c_n}]^i=X^i_{D^j_e(c_1),\dots,D^j_e(c_n)}.$$ 
Therefore, using  \eqref{e1}, we obtain
\begin{eqnarray*}
&&{\rm Lie}_e^l P^{(2)}_2={\rm Lie}_e^l d_{\omega_1}X_{c_1,\dots,c_n}=d_{\omega_1}{\rm Lie}_e^l X_{c_1,\dots,c_n}=d_{\omega_1}X_{D^l_e(c_1),\dots,D^l_e(c_n)}\ne 0\\
&&{\rm Lie}_e^{l+1} P^{(2)}_2={\rm Lie}_e^{l+1} d_{\omega_1}X_{c_1,\dots,c_n}=d_{\omega_1}{\rm Lie}_e^{l+1} X_{c_1,\dots,c_n}=d_{\omega_1}X_{D^{l+1}_e(c_1),\dots,D^{l+1}_e(c_n)}=0.
\end{eqnarray*} 
Now if \eqref{maxdeg} is satisfied, then $D^{l+1}_e(c_i)=0,\,\forall i$ and at least one of
 the functions $D^{l}_e(c_i)$ does not vanish. The
theorem follows from the observation that kernel of the map $(f_1,\dots,f_n)\to X_{f_1,\dots,f_n}$ is trivial.
\endproof

The next Theorem provides a first example on how it is possible to transfer information from $P^{(2)}_2$
 to all orders in $\epsilon$, if certain conditions are fulfilled.

\begin{theorem}\label{theoPoisson}
Let $\Pi_{\lambda}=\omega_{\lambda}+\sum_{k=1}^{\infty} \epsilon^{2k} P^{(2k)}_2$ 
be as in \eqref{poisson}, where $\omega_{\lambda}$ is an exact semisimple bi-Hamiltonian
 structure of hydrodynamic type and with $P^{(2)}_2$ in standard form.  Call $P_2:=\omega_2 +\sum_{k=1}^{\infty} \epsilon^{2k} P^{(2k)}_2$. If ${\rm Lie}^n_e P^{(2k)}_2=0$ for $k=1,\dots,N$ and if ${\rm Lie}^{n-1}_e P_2$ is a Poisson tensor for some $n\geq1$, $n$ integer, then there exists a Miura transformation $$\Pi_{\lambda}=\omega_{\lambda}+\sum_{k=1}^{\infty}\epsilon^{2k}
P^{(2k)}_2\to\tilde{\Pi}_{\lambda}
=\omega_{\lambda}+\sum_{k=1}^{N}\epsilon^{2k}
P^{(2k)}_2+\sum_{k=N+1}^{\infty}\epsilon^{2k}
\tilde{P}^{(2k)}_2$$
with
$${\rm Lie}^n_e \tilde{P}^{(2N+2)}_2=0.$$
\end{theorem}

Before proving Theorem \ref{theoPoisson}, let us observe the following. The condition that 
 ${\rm Lie}_e^{n-1}P_2$ is a Poisson tensor, in general,
 is not preserved by a Miura transformation.  This means
 that we cannot extend the result to higher order deformations. In Theorem \ref{theoPoisson}, for $n=1$ the extra requirement about ${\rm Lie}^{n-1}_e P_2$ being Poisson is automatically satisfied and is preserved by arbitrary Miura transformations ($P_2$ is indeed a Poisson tensor). In this form the Theorem has been proved in \cite{FL} and it states that if the central invariants are constants, then the {\em entire} pencil $\Pi_{\lambda}$ is exact, and not just its dispersionless limit. For $n\ne 1$ in order extend the
 theorem at any order we need to require, at each step, the additional assumption that ${\rm Lie}^{n-1}_e P_2$ is Poisson.
This is in general difficult to satisfy, although it is exactly what happens, for $n=2$, in the case of the single component
 Camassa-Holm equation (see next example). 

We will see later in Theorem \ref{fund} that it is possible to drop any additional condition on $P_2$ increasing as a function
 of $k$ the number of iterations of ${\rm Lie}_e$ acting on $P^{(2k)}_2$. 

\proof
Assume ${\rm Lie}^n_e P^{(2k)}_2=0$, for $k=1, \dots, N$. Than we show that  ${\rm Lie}^n_e \tilde{P}^{(2N+2)}_2=0$, where $  \tilde{P}^{(2N+2)}_2$ is obtained from $P^{(2N+2)}_2$ using a suitable Miura transformation. 
By assumption ${\rm Lie}^{n-1}_e P_2$ is Poisson, which means
 $[{\rm Lie}^{n-1}_e P_2, {\rm Lie}^{n-1}_e P_2]=0$ or equivalently
\begin{equation}
\label{lepoisson}d_{\omega_2} {\rm Lie}^{n-1}_e P^{(2l+2)}_2
=-\frac{1}{2}\sum_{k=1}^l\left[{\rm Lie}^{n-1}_e P^{(2k)}_2, {\rm Lie}^{n-1}_e P^{(2l-2k+2)}_2\right], 
\quad l\in \mathbb{N}.
\end{equation}
Taking \eqref{lepoisson} for $l=N$ and applying ${\rm Lie}_e$ on both sides, we get 
$${\rm Lie}_e d_{\omega_2} {\rm Lie}^{n-1}_e P^{(2N+2)}_2=-\frac{1}{2}\sum_{k=1}^N{\rm Lie}_e
\left(\left[{\rm Lie}^{n-1}_e P^{(2k)}_2, {\rm Lie}^{n-1}_e P^{(2N-2k+2)}_2\right]\right).$$
 By \eqref{e1}, \eqref{e2} and the fact that $d_{\omega_1} P^{(2k)}_2=0$ this last expression is equal to 
 $$d_{\omega_2} {\rm Lie}^n_e P^{(2N+2)}_2=$$
 $$=-\frac{1}{2}\sum_{k=1}^N\left(\left[{\rm Lie}^{n}_e P^{(2k)}_2, {\rm Lie}^{n-1}_e P^{(2N-2k+2)}_2\right]
+\left[{\rm Lie}^{n-1}_e P^{(2k)}_2, {\rm Lie}^{n}_e P^{(2N-2k+2)}_2\right]\right)=0.$$
 Moreover, by \eqref{e1} it is immediate to see that $d_{\omega_1} {\rm Lie}^n_e P^{(2N+2)}_2=0$.
 The fact that ${\rm Lie}^{n}_e P^{(2N+2)}_2$ is a cocycle for $d_{\omega_1}$ and $d_{\omega_2}$
 means that ${\rm Lie}^{n}_e P^{(2N+2)}_2={\rm Lie}_{X^{(2N+2)}_2}\omega_1$ for a suitable vector field $X^{(2N+2)}_2$ satisfying $d_{\omega_1} d_{\omega_2} (X^{(2N+2)}_2)=0$. 
Due to the triviality of
$H^2_{2N+2}(\mathcal{L}(\mathbb{R}^n),\omega_1,\omega_2)$ we have $$X^{(2N+2)}_2=d_{\omega_1} H^{(2N+2)}_2+d_{\omega_2} K^{(2N+2)}_2,$$
for suitable functionals $H^{(2N+2)}_2$ and $K^{(2N+2)}_2$ with densities that are differential polynomials. Now suppose to look for a functional $\tilde{K}^{(2N+2)}_2$ such that ${\rm Lie}^n_e \tilde{K}^{(2N+2)}_2=K^{(2N+2)}_2$.
 This equation with unknown $\tilde{K}^{(2N+2)}_2$ can always be solved and indeed it admits infinitely
 many solutions (see \cite{FL}). Consider now the Miura transformation generated by the vector field $\epsilon^{2N+2} d_{\omega_1} \tilde{K}^{(2N+2)}_2$. This transformation preserves $\omega_1$ since this vector field is a coboundary of $\omega_1$, while $P_2$ is changed in such a way that: 
$$\Pi_{\lambda}\rightarrow \tilde{\Pi}_{\lambda}=\omega_{\lambda}+\epsilon^2 P^{(2)}_2+\dots +\epsilon^{2N+2}\left(P^{(2N+2)}_2 +{\rm Lie}_{d_{\omega_1} \tilde{K}^{(2N+2)}_2}\omega_2\right)+O(\epsilon^{2N+4}).$$
Therefore $\tilde{P}^{(2N+2)}_2=\left(P^{(2N+2)}_2 +{\rm Lie}_{d_{\omega_1} \tilde{K}^{(2N+2)}_2}\omega_2\right)$. Now it is immediate to check that ${\rm Lie}^n_e\tilde{P}^{(2N+2)}_2=0$. Indeed, we have that
$${\rm Lie}^n_e\tilde{P}^{(2N+2)}_2={\rm Lie}^n_e{P}^{(2N+2)}_2+{\rm Lie}^n_e d_{\omega_2} d_{\omega_1}\tilde{K}^{(2N+2)}_2={\rm Lie}^n_e{P}^{(2N+2)}_2+d_{\omega_2}d_{\omega_1} K^{(2N+2)}_2,$$
due to \eqref{e1}, \eqref{e2} and ${\rm Lie}^n_e \tilde{K}^{(2N+2)}_2=K^{(2N+2)}_2$. Moreover, 
$$d_{\omega_2}d_{\omega_1} K^{(2N+2)}_2=-d_{\omega_1}d_{\omega_2}  K^{(2N+2)}_2=-{\rm Lie}_{X^{(2N+2)}_2}\omega_1=-{\rm Lie}^{n}_e P^{(2N+2)}_2$$ and thus ${\rm Lie}^n_e\tilde{P}^{(2N+2)}_2=0$. 
\endproof

\paragraph{{Camassa-Holm}}
Consider the Poisson pencil of the Camassa-Holm equation \cite{CH}:
$$2m\delta'(x-y)+m_x\delta(x-y)-\lambda(\delta'(x-y)-\delta'''(x-y)).$$
In the coordinate $u$ related to $m$ by the relation
$m=u+\epsilon u_x$
this pencil becomes
$$2u\delta'(x-y)+u_x\delta(x-y)-\lambda\delta'(x-y)+P_{2}$$
with\begin{eqnarray*}
P_{2}&=&\sum_{n=1}\epsilon^{2n}\left[\d_x\,u\,\delta^{(2n)}(x-y)+\d^{2n}_x\,u\,\delta'(x-y)\right]\\
&&+\sum_{n=1}\epsilon^{2n+1}\left(\d_x\left[u\delta^{(2n+1)}(x-y)\right]-\d_x^{2n+1}\left[u\delta'(x-y)\right]\right).
\end{eqnarray*}
It is easy to check that it satisfy the conditions.
\begin{eqnarray}
\label{ch1}
{\rm Lie}_e \omega_1&=&0\\
\label{ch2}
{\rm Lie}_e \omega_2&=&\omega_1\\
\label{wex}
{\rm Lie}^2_e P^{(2)}_2&=&0,
\end{eqnarray}
and moreover ${\rm Lie}_e P_2$  is Poisson, since ${\rm Lie}_e P_2$
 is a pencil with constant coefficients which is skew-symmetric. 

\paragraph{{Two component CH}}
In general the requirement that ${\rm Lie}_e^{n-1} P_2$
 is a Poisson tensor seems very restrictive.  
Let us consider for instance the Poisson pencil \eqref{PP2CH}.
Performing the Miura transformation
\begin{eqnarray*}
u&=&\xi\\
v&=&\eta+\epsilon \eta_x
\end{eqnarray*}
it becomes
\beq\label{PP2CHnew}
\tilde{P}_{\lambda}
=\begin{pmatrix}
          (2\xi\partial_x+\xi_x)\delta & (\eta+\epsilon
\eta_x)\sum_{k=0}^{\infty}\epsilon^k\delta^{(k+1)}\\
         
\sum_{k=0}^{\infty}(-1)^k\epsilon^k\d_x^{k+1}[(\eta+\epsilon
\eta_x)\delta]  &
-2\sum_{k=0}^{\infty}\epsilon^{2k}\delta^{(2k+1)}
         \end{pmatrix}-\lambda\begin{pmatrix}
          0 & \delta'\\ \delta' & 0
         \end{pmatrix}.
\eeq
The first order deformation can be written as
$$\begin{pmatrix}
          0 & \eta\delta''+\eta_x\delta'\\
          -\eta\delta''-\eta_x\delta'  & 0
         \end{pmatrix}=-{\rm Lie}_{X}\begin{pmatrix}
          (2\xi\d_x+\xi_x)\delta & \eta\delta'\\
          \d_x(\eta\delta)  & -2\delta'
         \end{pmatrix}$$
where
$$X=\begin{pmatrix}
          0 & \d_x\\ \d_x & 0
         \end{pmatrix}
\begin{pmatrix}
 \f{\delta H}{\delta\xi}\\
 \f{\delta H}{\delta\eta}
\end{pmatrix},\qquad H=-\int_{S^1}\f{\eta(x)^3}{12}\,dx.$$
This means that the Miura transformation generated by the
vector field $X$
 (up to terms of order $\mathcal{O}(\epsilon^3)$)
  reduces the pencil $\tilde{P}_{\lambda}$
 to the form $\tilde{\tilde{P}}_{\lambda}=$
\begin{eqnarray*}
&&\begin{pmatrix}
          (2\xi\d_x+\xi_x)\delta & \eta\delta'\\
          \d_x(\eta\delta)  & -2\delta'
         \end{pmatrix}-\lambda\begin{pmatrix}
          0 & \delta'\\
          \delta' & 0
         \end{pmatrix}+\\
&&\epsilon^2\left[\begin{pmatrix}
          0 & \eta\delta'''+\eta_x\delta''\\
          \eta\delta'''+2\eta_x\delta''+\eta_{xx}\delta' &
-2\delta'''
         \end{pmatrix}+\f{1}{2}{\rm
Lie}_{X}^2\begin{pmatrix}
          (2\xi\d_x+\xi_x)\delta & \eta\delta'\\
          \d_x(\eta\delta)  & -2\delta'
\end{pmatrix}\right]
\end{eqnarray*}
Since the dispersionless limit of the above pencil
coincides with the dispersionless limit of
 the pencil of the AKNS hierarchy, we have ${\rm
Lie}_Z\omega_2=\omega_1,\, {\rm Lie}_Z\omega_1=0$
 with $Z=\f{\d}{\d\eta}$. Moreover it is  easy to check 
that ${\rm Lie}_Z^3 P^{(2)}_2=0$ due to the
quadratic dependence of the central invariants on the
canonical coordinates. It is also possible to check that
the condition ${\rm Lie}_Z^3 P^{(n)}_2$ is no longer
satisfied by higher order deformations
 $(n>2)$ due to the presence of terms as 
$\eta^n\delta^{(n+1)}$.


\vspace{.5 cm}
\n
The above example shows that in some cases the assumption  ${\rm Lie}_e^n P^{(2)}_2=0$ is satisfied while
 the condition
  ${\rm Lie}_e^{n-1} P_2$ being Poisson is violated. In the remaining part of this section
 we will discuss what can be said  without this last condition. 

We have the following
\begin{theorem}\label{fund}
Let $\Pi_{\lambda}=\omega_{\lambda}+\sum_{k=1}^{\infty} \epsilon^{2k} P^{(2k)}_2$ 
be as in \eqref{poisson}, where $\omega_{\lambda}$ is an exact semisimple bi-Hamiltonian
 structure of hydrodynamic type and with $P^{(2)}_2$ in standard form. Call $P_2:=\omega_2 +\sum_{k=1}^{\infty} \epsilon^{2k} P^{(2k)}_2$. Suppose that ${\rm Lie}_e^n P_2^{(2)}=0$, for some positive integer $n$.
 Then there exists a Miura transformation that preserves $\omega_1$ and
 deforms $P_2$ to $\tilde{P}_2=\omega_2+\sum_{k=1}^{\infty}\epsilon^{2k}\tilde{P}^{(2k)}_2$ with
$${\rm Lie}^{nk-k+1}_e\tilde{P}^{(2k)}_2=0, \quad k=1, 2, \dots.$$
\end{theorem}
\proof
For $k=1$ there is nothing to prove, since this is the hypothesis of the Theorem. By inductive hypothesis we assume to have proved that we can construct at each step Miura transformations such that ${\rm Lie}^{kn-k+1}_e P^{(2k)}_2=0$ for $k=1, \dots, N$. (We call again the transformed components $P^{(2k)}_2$ although they have been obtained applying Miura transformations.) At the next step, we have ${\rm Lie}^{(N+1)(n-1)+1}_e P^{(2N+2)}_2\neq 0$ and we have to show that we can find a Miura transformation preserving $\omega_1$ such that the transformed term $\tilde{P}^{(2N+2)}_2$ satisfies ${\rm Lie}^{(N+1)(n-1)+1}_e\tilde{P}^{(2N+2)}_2=0$. 

First we show $d_{\omega_1}({\rm Lie}^{(N+1)(n-1)+1}_e P^{(2N+2)}_2)=0$. This is immediate since ${\rm Lie}_e$ commutes with $d_{\omega_1}$ and each $P^{(2k)}_2$ is a co-boundary of $\omega_1$ due to compatibility. Next we show $d_{\omega_2}({\rm Lie}^{(N+1)(n-1)+1}_e P^{(2N+2)}_2)=0.$ Using \eqref{e2} and the co-boundary property, we have that $d_{\omega_2}({\rm Lie}^{(N+1)(n-1)+1}_e P^{(2N+2)}_2)={\rm Lie}^{(N+1)(n-1)+1}_e(d_{\omega_2} P^{(2N+2)}_2)$. 
By the fact that $P_2$ is Poisson, we have $d_{\omega_2}P^{(2N+2)}_2=-\frac{1}{2}\sum_{k=1}^{N}[P^{(2k)}_2, P^{(2(N-k+1))}_2].$ 

It is therefore sufficient to show that ${\rm Lie}^{(N+1)(n-1)+1}_e[P^{(2k)}_2, P^{(2(N-k+1))}_2]=0$.
 For this, let us observe that 
\begin{eqnarray*}
&&{\rm Lie}^{(N+1)(n-1)+1}_e[P^{(2k)}_2, P^{(2(N-k+1))}_2]=\\
&&\sum_{j=0}^{(N+1)(n-1)+1}\binom{(N+1)(n-1)+1}{j}[{\rm Lie}^j_e P^{(2k)}_2, {\rm Lie}_e^{(N+1)(n-1)+1-j} P^{(2(N-k+1))}_2].
\end{eqnarray*}
By inductive hypothesis, ${\rm Lie}^j_e P^{(2k)}_2=0$ for $j\geq kn-k+1$, while 
$$ {\rm Lie}_e^{(N+1)(n-1)+1-j} P^{(2(N-k+1))}_2=0$$ for 
$(N+1)(n-1)+1-j\geq (N-k+1)n-(N-k+1)+1$ or equivalently for $j\leq kn-k$.
 Therefore each term is zero and we have proved that $d_{\omega_2}({\rm Lie}^{(N+1)(n-1)+1}_e P^{(2N+2)}_2)=0$.

Let us denote with $\alpha:=(N+1)(n-1)+1$. Since $d_{\omega_1} {\rm Lie}^{\alpha}_eP^{(2N+2)}_2=0$, $ {\rm Lie}^{\alpha}_eP^{(2N+2)}_2=d_{\omega_1} X^{(2N+2)}$, for a suitable vector field $ X^{(2N+2)}$. Moreover, since $ d_{\omega_2}{\rm Lie}^{\alpha}_eP^{(2N+2)}_2=0$ the vector field $X^{(2N+2)}$ defines a class in the bi-Hamiltonian cohomology group, which in this degree is however trivial. Thus $ X^{(2N+2)}=d_{\omega_1} H^{(2N+2)}_2+d_{\omega_2} K^{(2N+2)}_2$, for suitable local functionals $H^{(2N+2)}_2$ and $K^{(2N+2)}_2$. Since $ {\rm Lie}^{\alpha}_eP^{(2N+2)}_2=d_{\omega_1} X^{(2N+2)}$, we also have \begin{equation}\label{eq1}  {\rm Lie}^{\alpha}_eP^{(2N+2)}_2=d_{\omega_1}d_{\omega_2} K^{(2N+2)}_2.\end{equation}

To construct the Miura transformation we are looking for, we proceed as follows. Given the functional $K^{(2N+2)}_2$, we can always find another functional $\tilde{K}^{(2N+2)}_2$ such that ${\rm Lie}^{\alpha}_e\tilde{K}^{(2N+2)}_2=K^{(2N+2)}_2$. Indeed, it has been proved in \cite{FL} that the equation ${\rm Lie}_e K_1=K^{(2N+2)}_2$ is always solvable given $K^{(2N+2)}_2$, and therefore, repeating with ${\rm Lie}_e K_2 =K_1$, etc. one gets that there exists a local functional $\tilde{K}^{(2N+2)}_2$ with ${\rm Lie}^{\alpha}_e \tilde{K}^{(2N+2)}_2=K^{(2N+2)}_2$. 

Now we consider the Miura transformation generated by the vector field $\epsilon^{2N+2}d_{\omega_1}\tilde{K}^{(2N+2)}_2$. Since this is a co-boundary of $d_{\omega_1}$, $\omega_1$ is undeformed, while 
$$P_2\mapsto \tilde P_2=\omega_2 +\sum_{k=1}^{N}\epsilon^{2k} P^{(2k)}_2+\epsilon^{2N+2}(\tilde{P}^{(2N+2)}_2)+O(\epsilon^{(2N+4)}),$$
where $$\tilde{P}^{(2N+2)}_2=P^{(2N+2)}_2+{\rm Lie}_{d_1\tilde{K}^{(2N+2)}_2}\omega_2.$$
Therefore 
$${\rm Lie}^{\alpha}_e\tilde{P}^{(2N+2)}_2={\rm Lie}_e^{\alpha}P^{(2N+2)}_2+{\rm Lie}_e^{\alpha}(d_{\omega_2} d_{\omega_1} \tilde{K}^{(2N+2)}_2)=$$
$$=d_{\omega_1}d_{\omega_2}K^{(2N+2)}_2+d_{\omega_2}d_{\omega_1}K^{(2N+2)}_2=0,$$
the first term by equation \eqref{eq1} and the second due to the fact that ${\rm Lie}^{\alpha}_e$ commutes with $d_{\omega_2}d_{\omega_1}$. 
\endproof

Proposition \ref{centralinvariantsn} and theorem \ref{fund} can be summarized in the following
 theorem
\begin{theorem}\label{theoPoisson2}
Let $\Pi_{\lambda}$ a Poisson pencil with
 polynomial central invariants of maximal degree $n-1$ and suppose
 that its dispersionless limit $\omega_{\lambda}$ is exact.
 Then  there exists a Miura transformation reducing the pencil to the form $\tilde\Pi_{\lambda}=\omega_{\lambda}+\sum_{k=1}^{\infty}\epsilon^{2k}\tilde{P}^{(2k)}_2$
 with
$${\rm Lie}^{nk-k+1}_e\tilde{P}^{(2k)}_2=0,
 \quad k=1, 2, \dots.$$
\end{theorem}

In view of Theorem \ref{theoPoisson2} we introduce the following 
\begin{definition}
Let $\omega_1$ and $\omega_2$ a pair of exact Poisson structures with ${\rm Lie}_e \omega_2=\omega_1$ and ${\rm Lie}_e\omega_1=0$.
 Consider the compatible pencil 
$\omega_1$ and $P_2:=\omega_2+\sum_{k=1}^{\infty} \epsilon^{2k} P^{2k}_2$, with $P^{(2)}_2$ in standard form and
 ${\rm Lie}^n_e P^{(2)}_2=0$.
 Then the pair $(\omega_1, P_2)$ is said to be in {\em normal form} if  ${\rm Lie}^{nk-k+1}_e P^{(2k)}_2=0$ for all $k\in \mathbb{N}$. 
\end{definition}

Theorem  \ref{theoPoisson2} can be reformulated by saying that any Poisson pencil $\Pi_{\lambda}$ with {\em polynomial} central invariants whose dispersionless limit is an exact Poisson pencil of hydrodynamic type can be put in normal form via a sequence of Miura transformations. 

Using these tools we can map a Poisson pencil with central invariants that are polynomials of degree
 $n-1$ to a pencil with constant central invariants. This map is somehow mysterious
 since at this point we do not have an interpretation of it in terms of known transformations, 
like a combination of Miura and reciprocal transformations.

\begin{theorem}
Let $(\omega_2, \omega_1)$ be an exact bi-Hamiltonian structure with respect to the Liouville vector field $e$ (${\rm Lie}_e\omega_2=\omega_1$, ${\rm Lie}_e\omega_1=0$), such that $\omega_1$ and $P_2=\omega_2+\sum_{k=1}^{\infty} \epsilon^{2k}P^{(2k)}_2$ are compatible Poisson structures. Furthermore, assume ${\rm Lie}^n_eP^{(2)}_2=0$, ${\rm Lie}^{n-1}_e P^{(2)}_2\neq0$ and that the pair $(\omega_1, P_2)$ is in normal form. Then  the pair of compatible Poisson structures $\omega_1$ and $Q_2:=\omega_2+\sum_{k=1}^n Q^{(2k)}_2$, where $$Q^{(2k)}_2:=\f{{\rm Lie}^{nk-k}_e P^{(2k)}_2}{[k(n-1)]!}$$
  has constant central invariants and it is also again in normal form.  
\end{theorem}
\proof
First of all, we notice that ${\rm Lie}_e Q^{(2)}_2=0$, so by \cite{FL} the pair $\omega_1$ and $Q_2$
 has constant central invariants.
Now it remains to prove that the pair $\omega_1$ and $Q_2$ is a pair of compatible Poisson structures.
 The compatibility of $Q_2$ with $\omega_1$ is immediate, we need to prove that $Q_2$ is Poisson.
 For this we show that 
\begin{equation}\label{eq3}
d_{\omega_2} Q^{(2N+2)}_2=-\frac{1}{2}\sum_{k=1}^N\left[Q^{(2k)}_2, Q^{(2(N-k+1))}_2\right].
\end{equation}
Substituting everywhere in \eqref{eq3} the expression for $Q^{(2l)}_2={\rm Lie}^{nl-l}_eP^{(2l)}_2$ we find
$$d_{\omega_2} {\rm Lie}^{(N+1)(n-1)}_e P^{(2(N+1))}_2=-\frac{1}{2}\sum_{k=1}^N\left[
{\rm Lie}^{k(n-1)}_e P^{(2k)}_2, {\rm Lie}^{(N-k+1)(n-1)}_e P^{(2(N-k+1))}_2\right].$$
Since each $P^{(2l)}_2$ is a $d_{\omega_1}$ co-boundary, we have we that $d_{\omega_2} {\rm Lie}^{(N+1)(n-1)}_e P^{(2(N+1))}_2$ is equal to $ {\rm Lie}^{(N+1)(n-1)}_e d_{\omega_2} P^{(2(N+1))}_2$.
Since $P_2$ is Poisson, we have  
\begin{equation}\label{eq6} {\rm Lie}^{(N+1)(n-1)}_e d_{\omega_2} P^{(2(N+1))}_2
= {\rm Lie}^{(N+1)(n-1)}_e \left(-\frac{1}{2}\sum_{k=1}^N\left[P^{(2k)}_2, P^{(2(N-k+1))}_2\right]\right). 
\end{equation}
On the other hand 
\begin{eqnarray}
\nonumber
&&{\rm Lie}^{(N+1)(n-1)}_e\left[P^{(2k)}_2, P^{(2(N-k+1))}_2\right]=\\
\label{eq4}
&&\sum_{j=0}^{(N+1)(n-1)}\binom{(N+1)(n-1)}{j}\left[{\rm Lie}^j_e P^{(2k)}_2, {\rm Lie}^{(N+1)(n-1)-j}_e
P^{(2(N-k+1))}_2\right],
\end{eqnarray}
and since $(\omega_1, P_2)$ is in normal form, we have that ${\rm Lie}^j_e P^{(2k)}_2=0$ for $j\geq nk-k+1$ and ${\rm Lie}^{(N+1)(n-1)-j}_eP^{(2(N-k+1))}_2=0$ for 
$(N+1)(n-1)-j\geq n(N-k+1)-(N-k+1)+1$ or equivalently for $j\leq nk-k-1$.
 Therefore, the only surviving term in the sum on the right hand side
 of equation \eqref{eq4} corresponds to $j=nk-k$. Thus we have 
\begin{eqnarray}
\nonumber
&&{\rm Lie}^{(N+1)(n-1)}_e\left[P^{(2k)}_2, P^{(2(N-k+1))}_2\right]=\\
\label{eq5}
&&\binom{(N+1)(n-1)}{k(n-1)}\left[{\rm Lie}^{nk-k}_e P^{(2k)}_2, {\rm Lie}^{(N-k+1)(n-1)}_e 
P^{(2(N-k+1))}_2\right].
\end{eqnarray}
Substituting \eqref{eq5} in the right hand side of \eqref{eq6} we get
\begin{eqnarray}
\nonumber
&&{\rm Lie}^{(N+1)(n-1)}_e d_{\omega_2} P^{(2(N+1))}_2=\\
&&-\frac{1}{2}\sum_{k=1}^N \binom{(N+1)(n-1)}{k(n-1)}\left[{\rm Lie}^{nk-k}_e P^{(2k)}_2,
 {\rm Lie}^{(N-k+1)(n-1)}_e P^{(2(N-k+1))}_2\right],
\end{eqnarray}
or
\beq
d_{\omega_2}\left(\f{{\rm Lie}^{(N+1)(n-1)}_e  P^{(2(N+1))}_2}{[(N+1)(n-1)]!}\right)=
-\frac{1}{2}\sum_{k=1}^N\left[\f{{\rm Lie}^{nk-k}_e P^{(2k)}_2}{[k(n-1)]!}, \f{{\rm Lie}^{(N-k+1)(n-1)}_e P^{(2(N-k+1))}_2}{[(N+1-k)(n-1)]!}\right],
\eeq
that is
$$d_{\omega_2}Q^{(2(N+1))}_2=
-\frac{1}{2}\sum_{k=1}^N\left[Q^{(2k)}_2, Q^{(2(N-k+1))}_2\right].$$
This proves that $Q_2$ is indeed Poisson and thus we have transformed a pencil in normal form $(\omega_1, P_2)$
 with central invariants of order $n-1$ to a pencil $(\omega_1, Q_2)$ with constant central invariants.
 Notice also that $(\omega_1, Q_2)$ is still in normal form: in this case ${\rm Lie}_e Q^{(2)}_2=0$
 and also ${\rm Lie}_e Q^{(2k)}_2=0$, as it is immediate to check.  
\endproof

\paragraph{{The case of $r$-KdV-CH hierarchy}}
As an application of the previous Theorem, 
let us study how ${\rm Lie}^r_e$ acts on the $r$-KdV-CH hierarchy. We will use the notation introduced
 in Section 3. We will show the following
\begin{proposition}
In the case $k=0$, $l=1$, i.e. in the case in which the central invariants are polynomials of degree $r$ parametrized by $(a_0\;, \;\dots\;,\;a_r)$, ${\rm Lie}^r_e$ acts on the $r$-KdV-CH hierarchy, mapping the system with parameters $(a_0\; ,\; \dots \; ,\; a_r)$ to the one with parameters $(a_r \; ,\; 0\;, \; \dots\; ,\;  0)$.
\end{proposition}
\proof 
We fix our attention to the bi-Hamiltonian structure $(P_k, P_l)$, where $k$ and $l$ are general for the moment.
 First, in $\lambda_i$ coordinates, we normalize $P_k$ up to the order $O(\epsilon^4)$,
 eliminating with a Miura transformation the term in $\epsilon^2$. In particular, 
 looking the leading term in derivatives of  $\delta(x-y)$ in $\epsilon^2$, 
one can use as an ansatz for the vector field $Z$ generating the Miura transformation 
$$Z^i=\epsilon^2(b^i(\lambda)\lambda_{i,xx}+\dots)+O(\epsilon^4).$$
 Now it is immediate to see that 
 $${\rm Lie}_Z P^{ii}_k=-\epsilon^2\left[ b^i(\lambda)g^{ii}_k \delta^{(3)}(x-y)+\dots\right].$$
 Since we want to normalize $P^{ij}_k$ up to $O(\epsilon^4)$, we have that we have to impose
 $${\rm Lie}_Z P^{ii}_k=-\epsilon^2\left(E^{ii}_k(\lambda)+\dots \right)=-\epsilon^2\left[\frac{\lambda^k_i\frak{a}'(\lambda_i)-k\lambda^{k-1}_i\frak{a}(\lambda_i)}{2(P'(\lambda_i))^2} \right]\delta^{(3)}(x-y)+\dots.$$
 Since $g^{ii}_k=-\frac{2\lambda^k_i}{P'(\lambda_i)},$ one gets
 $$b_i(\lambda_i)=-\frac{\frak{a}'(\lambda_i)-k\frak{a}(\lambda_i)/\lambda_i}{4P'(\lambda_i)}.$$
 In this way, we eliminate the term in $\epsilon^2$ from $P_k$. Now the same Miura transformation applied to $P_l$ is computed as follows. 
 Using the formula for Lie derivatives one finds:
 $${\rm Lie}_Z P^{ii}_l=-\epsilon^2\left(b_i(\lambda_i)g^{ii}_l \delta^{(3)}(x-y)+\dots\right)$$
 and substituting the expression for $b_i(\lambda_i)$ computed above and the expression for $g^{ii}_l$ we find
 $${\rm Lie}_Z P^{ii}_l=\epsilon^2\left[-\frac{\frak{a}'(\lambda_i)-k\frak{a}(\lambda_i)/\lambda_i}{4P'(\lambda_i)} \frac{2\lambda^l_i}{P'(\lambda_i)}\delta^{(3)}(x-y)+\dots \right].$$
 Therefore $P^{ii}_l$ is mapped to 
 $$P^{ii}_l\mapsto (g^{ii}_l\delta^{(1)}(x-y)+\dots)+\epsilon^2\left[ \left(E^{ii}_l(\lambda)-\frac{\frak{a}'(\lambda_i)-k\frak{a}(\lambda_i)/\lambda_i}{4P'(\lambda_i)} \frac{2\lambda^l_i}{P'(\lambda_i)}\right)\delta^{(3)}(x-y)+\dots \right],$$
 so the term in $\epsilon^2$ in $P^{ii}_l$ is reduced to 
 $$\epsilon^2\left[\frac{(k-l)\lambda^{l-1}_i \frak{a}(\lambda_i)}{2(P'(\lambda_i))^2}\delta^{(3)}(x-y)+\dots \right].$$
 Now we specialize to $k=0$, so that the exactness condition $\sum_h \frac{\partial g^{ii}_k}{\partial u^h}=0\; $, $\; i=0, \dots, r-1$ holds, and furthermore we assume $l=1$ so that the central invariants are polynomials of degree $r$. Now observe that in the case $k=0$, $l=1$, the canonical coordinates $u^i=\lambda_i$, so the vector field $e=\sum_h \frac{\partial}{\partial u^h}=\sum_h \frac{\partial}{\partial \lambda_h}.$ Let us observe also that $\sum_h \frac{\partial}{\partial \lambda_h}P'(\lambda_i)=0.$ Taking into account these two facts and applying $\f{1}{r!}{\rm Lie}^r_e$ to the term $\epsilon^2$ in $P^{ij}_{l=1}$ we obtain 
 $$ P^{ij}_{l=1}=(g^{ij}_1\delta^{(1)}(x-y)+\dots)+\epsilon^2\left(-\frac{a_r}{2(P'(\lambda_i))^2}\delta^{(3)}(x-y)+\dots\right),$$
 that is, using ${\rm Lie}^r_e$ we can map an $r$-KdV-CH system associated to $\frak{a}(\lambda)$ to an $r$-KdV-CH system associated to $\f{1}{r!}\frac{d^r}{d\lambda^r}\frak{a}(\lambda).$
\endproof

\section{Deformations of homogeneous Poisson pencils
 of hydrodynamic type}\label{sechomogeneous}
Let us consider a deformation $P_{\lambda}$ of a homogenous Poisson pencil of hydrodynamic type $\omega_{\lambda}$. We will call it
 \emph{homogenous} if its central invariants are homogenous functions of the \emph{same} degree $D$ in the canonical coordinates:
$$\sum_{k=1}^n u^k \f{\d c_i}{\d u^k}=D c_i.$$
As usual we can assume, without loss of generality, that  the pencil $P_{\lambda}$ has the form
 $$P_{\lambda}=\omega_2+\sum_{k=1}^{\infty}
\epsilon^{2k}P^{(2k)}_2-\lambda\omega_1,$$
with $P^{(2)}_2$ in standard form. 
In this case we know that
$$P^{(2)}_2=d_{\omega_1}d_{\omega_2}\left(\sum_{i=1}^N
\int c^i(u^i)u^i_x\ln{u^i_x}\,dx\right).$$
Using the properties  \eqref{E1} and  \eqref{E2} and the identity
$${\rm Lie}_E \sum_{i=1}^N
\int c^i(u^i)u^i_x\ln{u^i_x}\,dx=(D+1)\sum_{i=1}^N
\int c^i(u^i)u^i_x\ln{u^i_x}\,dx$$
we obtain
\beq\label{Hom3}
{\rm Lie}_E P^{(2)}_2=\left[2(d-1)+D\right]P^{(2)}_2
\eeq
Indeed we have
\begin{eqnarray*}
&&{\rm Lie}_E P^{(2)}_2={\rm Lie}_E d_{\omega_1}d_{\omega_2}\left(\sum_{i=1}^N
\int c^i(u^i)u^i_x\ln{u^i_x}\,dx\right)=\\
&&d_{\omega_1}d_{\omega_2}{\rm Lie}_E\left(\sum_{i=1}^N
\int c^i(u^i)u^i_x\ln{u^i_x}\,dx\right)+(2d-3)d_{\omega_1}d_{\omega_2}\left(\sum_{i=1}^N
\int c^i(u^i)u^i_x\ln{u^i_x}\,dx\right)=\\
&&(D+1+2d-3)d_{\omega_1}d_{\omega_2}\left(\sum_{i=1}^N
\int c^i(u^i)u^i_x\ln{u^i_x}\,dx\right)=\left[2(d-1)+D\right]P^{(2)}_2.
\end{eqnarray*}

The behaviour of the the flat pencil
 $g_{\lambda}$ with respect to the Euler vector field
 has some consequences also on the form of the higher
 order deformations. We have the following theorem
\begin{theorem}
A homogeneous Poisson pencil $P_\lambda$
 with homogeneous central invariants of degree $D$ can be reduced by a Miura transformation to a Poisson pencil
 $Q_\lambda$ of the form
\beq\label{HomPencil}
Q_{\lambda}=\omega_2+\sum_{k=1}^{\infty}
\epsilon^{2k}Q^{(2k)}_2-\lambda\omega_1
\eeq
 with
\beq\label{homPP}
{\rm Lie}_E Q_{2}^{(2k)}=[(k+1)(d-1)+kD]Q_{2}^{(2k)},\qquad k=1,2,\dots
\eeq
\end{theorem}

\n
\emph{Proof}. We can prove the theorem by induction. Suppose that 
$${\rm Lie}_E P_{2}^{(2k)}=[(k+1)(d-1)+kD]P_{2}^{(2k)},\qquad k=1,\dots,N$$
and that
$${\rm Lie}_E P_{2}^{(2N+2)}\ne[(N+2)(d-1)+(N+1)D]P_{2}^{(2N+2)},\qquad k=1,\dots,N$$
then we show that there exists a Miura transformation, canonical
 with respect to $\omega_1$, reducing the pencil
 $P_{\lambda}$ to the form
\beq\label{HomPencilN}
Q_{\lambda}=\omega_2+\sum_{k=1}^{\infty}
\epsilon^{2k}Q^{(2k)}_2-\lambda\omega_1
\eeq
 with
\beq\label{homPPN}
{\rm Lie}_E Q_{2}^{(2N+2)}=[(N+2)(d-1)+(N+1)D]Q_{2}^{(2N+2)}.
\eeq

We need to show that
\begin{eqnarray*}
d_{\omega_1}
\left(
{\rm Lie}_E P_{2}^{(2N+2)}-[(N+2)(d-1)+(N+1)D]P_{2}^{(2N+2)}
\right)&=&0\\
d_{\omega_2}
\left(
{\rm Lie}_E P_{2}^{(2N+2)}-[(N+2)(d-1)+(N+1)D]
P_{2}^{(2N+2)}
\right)&=&0
\end{eqnarray*}
The first property follows immediately \eqref{E1} and from $d_{\omega_1}P_{2}^{(2k)}=0$ (which follows from the compatiblity
 of the pencil $P_{\lambda}$), while
 the second one can be proved using the induction hypothesis. Indeed, taking into account that
$$d_{\omega_2}P_{2}^{(2N+2)}=-\f{1}{2}\sum_{k=1}^N\left[P_{2}^{(2k)},P_{2}^{(2N+2-2k)}\right]$$
we obtain
\begin{eqnarray*}
&&d_{\omega_2}\left(
{\rm Lie}_E P_{2}^{(2N+2)}-[(N+2)(d-1)+(N+1)D]
P_{2}^{(2N+2)}\right)=\\
&&{\rm Lie}_E d_{\omega_2}P_{2}^{(2N+2)}
-(d-1)d_{\omega_2}P_{2}^{(2N+2)}-[(N+2)(d-1)+(N+1)D]
d_{\omega_2}P_{2}^{(2N+2)}=\\
&&-\f{1}{2}{\rm Lie}_E\sum_{k=1}^N\left[P_{2}^{(2k)},P_{2}^{(2N+2-2k)}\right]-[(N+3)(d-1)+(N+1)D]
d_{\omega_2}P_{2}^{(2N+2)}=\\
&&-\sum_{k=1}^N\left[{\rm Lie}_E P_{2}^{(2k)},P_{2}^{(2N+2-2k)}\right]-[(N+3)(d-1)+(N+1)D]
d_{\omega_2}P_{2}^{(2N+2)}=\\
&&-\sum_{k=1}^N[(k+1)
(d-1)+kD]\left[P_{2}^{(2k)},P_{2}^{(2N+2-2k)}\right]-[(N+3)(d-1)+(N+1)D]
d_{\omega_2}P_{2}^{(2N+2)}
\end{eqnarray*}
Since $[P_{2}^{(2k')},P_{2}^{(2N+2-2k')}]=[P_{2}^{(2N+2-2k')},P_{2}^{(2k')}]$, we
 can reorganize the sum in such a way that the terms with $k=k'$ and $k=N+1-k'$ have the \emph{same} coefficient
\begin{eqnarray*}
&&\f{1}{2}\left\{[(k'+1)(d-1)+k'D]+[(N+2-k')(d-1)+(N+1-k')D]\right\}=\\
&&\f{1}{2}[(N+3)(d-1)+(N+1)D],
\end{eqnarray*}
 \emph{independent} on $k'$. The result is 
\begin{eqnarray*}
&&-\sum_{k=1}^N[(k+1)
(d-1)+kD]\left[P_{2}^{(2k)},P_{2}^{(2N+2-2k)}\right]-[(N+3)(d-1)+(N+1)D]
d_{\omega_2}P_{2}^{(2N+2)}\\
&&[(N+3)(d-1)+(N+1)D]\left(-\f{1}{2}\sum_{k=1}^N
\left[P_{2}^{(2k)},P_{2}^{(2N+2-2k)}\right]-
d_{\omega_2}P_{2}^{(2N+2)}\right)=0.
\end{eqnarray*}
According to the results of Liu and Zhang there exists a vector  field 
$$X_2^{(2N+2)}=d_{\omega_1} H_2^{(2N+2)}+d_{\omega_2} K_2^{(2N+2)}$$
such that
$${\rm Lie}_E P_{2}^{(2N+2)}-[(N+2)(d-1)+(N+1)D]
P_{2}^{(2N+2)}=d_{\omega_1}X^{(2N+2)}
=d_{\omega_1}d_{\omega_2}K_2^{(2N+2)}.$$
The Miura transformation generated by the vector field
 $d_{\omega_1}\tilde{K}_2^{(2N+2)}$ reduces the pencil
 to the form $\omega_2+\sum_k\epsilon^{2k}Q_2^{(2k)}$ where
\begin{eqnarray*}
Q_2^{(2k)}&=&P_2^{(2k)},\qquad k=1,\dots,N\\
P_2^{(2N+2)}&=&P_2^{(2N+2)}-d_{\omega_1}d_{\omega_2}\tilde{K}_2^{(2N+2)}.
\end{eqnarray*}
Moreover the ``new'' pencil satisfies
\begin{eqnarray*}
&&{\rm Lie}_E Q_2^{(2N+2)}={\rm Lie}_E(P_2^{(2N+2)}-d_{\omega_1}d_{\omega_2}\tilde{K}_2^{(2N+2)})=\\
&&d_{\omega_1}d_{\omega_2}K_2^{(2N+2)}+[(N+2)(d-1)+(N+1)D]P_{2}^{(2N+2)}-{\rm Lie}_E d_{\omega_1}
d_{\omega_2}\tilde{K}_2^{(2N+2)}=\\
&&d_{\omega_1}d_{\omega_2}K_2^{(2N+2)}+[(N+2)(d-1)+(N+1)D]P_{2}^{(2N+2)}+\\
&&-d_{\omega_1} d_{\omega_2}\left({\rm  Lie}_E\tilde{K}_2^{(2N+2)}+(2d-3)
\tilde{K}_2^{(2N+2)}\right).
\end{eqnarray*}
If the functional $\tilde{K}_2^{(2N+2)}$ satisfies the further condition
$${\rm  Lie}_E\tilde{K}_2^{(2N+2)}+(2d-3)
\tilde{K}_2^{(2N+2)}-K_2^{(2N+2)}=[(N+2)(d-1)+(N+1)D]\tilde{K}_2^{(2N+2)}$$
that is
\beq\label{qlpde}
{\rm  Lie}_E\tilde{K}_2^{(2N+2)}-[N(d-1)+(N+1)D+1]
\tilde{K}_2^{(2N+2)}=K_2^{(2N+2)}
\eeq
then
$${\rm Lie}_E Q_2^{(2N+2)}=
[(N+2)(d-1)+(N+1)D]Q_2^{(2N+2)}$$
as required.

To conclude the proof we have to show that an equation of the form
\beq\label{qlpde2}
{\rm  Lie}_E\tilde{K}+C\tilde{K}=K
\eeq
always admits solutions (for $C=0$ it has been proved in \cite{FL}). In canonical coordinates \eqref{qlpde2} reads
$$\sum_{i=1}^n\int_{S^1}\left[u^i\f{\d\tilde{k}}{\d u^i}+\sum_{s=1}^{\infty}u^i_{(s)}\f{\d\tilde{k}}{\d u^i_{(s)}}\right]\,dx
=\int_{S^1}k\,dx$$
A solution can be found solving the equation
\beq\label{eqdens}
\sum_{i=1}^n\left[u^i\f{\d\tilde{k}}{\d u^i}+\sum_{s=1}^{\infty}u^i_{(s)}\f{\d\tilde{k}}{\d u^i_{(s)}}\right]=k
\eeq
for the density of the functional $\tilde{K}$. The differential operator 
$$\sum_{s=1}^{\infty}u^i_{(s)}\f{\d}{\d u^i_{(s)}}$$
acts on the single monomials of $\tilde{K}$ multiplying them by the their degree as polinomials in the
 variables $u^i_x,\,u^i_{xx},\dots, u^i_{(s)},\dots$.
 Notice that this is different from the degree they have as differential polinomials. In the last case
 ${\rm deg}(u^i_{(s)})=s$ while in the former case ${\rm deg}(u^i_{(s)})=1$. 
This means that the equation \eqref{eqdens} 
 is equivalent to the quasilinear equations 
$$\sum_{i=1}^n u^i\f{\d\tilde{A_j}}{\d u^i}+c_j\tilde{A_j}=A_j,\qquad\sum_{i=1}^n u^i\f{\d\tilde{B}_{jm}}{\d u^i}+c_{jm}\tilde{B}_{jm}=B_{jm},\qquad\dots$$
for the coefficients $\tilde{A}_i,\tilde{B}_{ij},\, {
\rm etc.}$ of the homogenous differential 
 polynomial 
$$\tilde{k}=\tilde{A}_i u^i_{(N)}+\tilde{B}_{ij} u^i_x u^j_{(N-1)}+\dots.$$
The constant $c_j,\,c_{jm},\dots$ are equal to $C$ plus the degree of the monomial containing $A_j,\,B_{jm},$ and so on. For instance $c_j=C+1$,
 $c_{jm}=C+2$, etc.
It is well known (see for instance \cite{CouHil} or \cite {Sm})
 that equations of this form admit $n$ functional independent solutions.

 Indeed plugging-in $u^{n+1}=\tilde{A}$ and looking for solutions of
$$\sum_{i=1}^n u^i\f{\d u^{n+1}}{\d u^i}+cu_{n+1}=A(u^1,\dots,u^n)$$
in implicit form:
$$\phi(u^1,\dots,u^{n+1})={\rm cost}$$
we obtain
$$-\sum_{i=1}^n u^i\f{\f{\d\phi}{\d u^i}}{\f{\d\phi}{\d u^{n+1}}}=A(u^1,\dots,u^n)-cu^{n+1}$$
that is
$$\sum_{i=1}^n u^i\f{\d\phi}{\d u^i}+(A(u^1,\dots,u^n)-cu^{n+1})\f{\d\phi}{\d u^{n+1}}$$
which is the quasilinear equation for the first integrals of the vector field
$$\sum_{i=1}^n\f{\d}{\d u^i}+(A(u^1,\dots,u^n)-cu^{n+1})\f{\d}{\d u^{n+1}}.$$ 
\endproof

An interesting consequence of the previous Theorem is that the tensorial quantities appearing in the $
\epsilon$-expansion of the Poisson pencil have a well-defined degree of homogeneity with respect to $E$. 
This is detailed in the following:  
\begin{corollary}
The tensor fields $A^{ij}_{(2)k,0},\,k=2,4,6,\dots$ appearing in the leading terms of the $\epsilon$-expansion of the
 Poisson pencil \eqref{HomPencil}
\begin{eqnarray*}
Q^{ij}_{\lambda}=\omega^{ij}_{2}+\sum_{k\ge 1}\epsilon^{2k}\left\{\sum_{l=0}^{2k+1}A^{ij}_{(2)2k,l}(q,q_x,\dots,q_{(l)})\delta^{(2k-l+1)}(x-y)+\dots\right\}
-\lambda\omega^{ij}_{1}
\end{eqnarray*}
satisfy the homogeneity condition
\beq\label{homCond}
{\rm Lie}_E A_{(2)2k,0}=[(k+1)(d-1)+kD]A^{ij}_{(2)2k,0},\qquad k=1,\dots,N.
\eeq
Here we identify the Euler vector field on the manifold with the corresponding vector field on the loop space.
\end{corollary}

\n
\proof The proof is a direct consequence of the formula
\begin{eqnarray}\label{Lie}
&&{\rm Lie}_{E}Q_2^{ij}=\\
&&\sum_{k,s}\left( \partial^s_x E^k(u(x), \dots) 
\frac{\partial Q_2^{ij}}{\partial u^k_{(s)}(x)}
- \frac{\partial E^i(u(x), \dots)}{\partial u^k_{(s)}(x)} \partial ^s_x Q_2^{kj}
-\frac{\partial E^j(u(y), \dots)}{\partial u^k_{(s)}(y)}\partial^s_y Q_2^{ik}\right)\nn,
\end{eqnarray}
Indeed, considering only the leading terms in the $\epsilon$-expansion equation \eqref{homPP} reads
\beq\label{homPP2}
\left[{\rm Lie}_E A^{ij}_{(2)2k,0}\delta^{(2k+1)}(x-y)+\dots\right]=
[(k+1)(d-1)+kD]\left[A^{ij}_{(2)2k,0}\delta^{(2k+1)}(x-y)+\dots\right]
\eeq
\endproof

As an example, we check the homogeneity predictions of the previous Theorem in the case of non trivial bi-Hamiltonian deformations of the Poisson pencil 
 $\omega_{\lambda}=\omega_2+\lambda \omega_1=u\delta'(x-y)+\f{1}{2}u_x\delta(x-y)+\lambda\delta'(x-y)$, up to order $\epsilon^6$. 
 We will need the following auxiliary formula
  \begin{equation}\label{auxiliary}
 \frac{\partial}{\partial u_{(s)}}\partial^l_x=\sum_{t=0}^l\binom{l}{t}\partial^t_x \frac{\partial }{\partial u_{(s-l+t)}},
 \end{equation}
 and the following simple 
 \begin{lemma} The following identity holds
 \begin{equation}\label{auxiliary2}
 \sum^l_{s= 0}(\partial^s_x u)\frac{\partial }{\partial u_{(s)}}(\partial^l_x u^D)=D\partial^l_x u^D.
 \end{equation}
 \end{lemma}
 \proof 
 By \eqref{auxiliary} we can write the left hand side of \eqref{auxiliary2} as
 $$\sum^l_{s= 0}(\partial^s_x u)\frac{\partial }{\partial u_{(s)}}(\partial^l_x u^D)=\sum^l_{s= 0}(\partial^s_x u)\sum_{t=0}^l \binom{l}{t} \partial^t_x\left(\frac{\partial}{\partial u_{(s-l+t)}} u^D \right)=$$
 $$=D\sum_{s=0}^l (\partial^s_x u) \binom{l}{l-s}\partial^{l-s}_x u^{D-1}=D\partial^l_x u^D.$$
 \endproof
 The non trivial bi-Hamiltonian deformations of $\omega_{\lambda}$ up the $\epsilon^6$ have been computed in \cite{AL} (extending
 previous results of \cite{L}) providing the following result:
 \begin{theorem}
Up  to Miura transformations, the deformations of the pencil $\omega_{\lambda}=u\delta^{(1)}(x-y)+\frac{1}{2}u_{(1)}\delta(x-y)-\lambda \delta^{(1)}(x-y)$
can be reduced to the following form:
\begin{equation}\label{main.eq}
\begin{split}
Q_{\lambda}=\omega_{\lambda}-\epsilon^2\left\{\partial^2_x\left(c_2\delta^{(1)}(x-y)\right)+c_2 \delta^{(3)}(x-y)+(\partial_x c_2)\delta^{(2)}(x-y)\right\}\\
-\epsilon^4 \left\{\partial^4_x \left(c_4 \delta^{(1)}(x-y)\right)+c_4\delta^{(5)}(x-y)+(\partial_x c_4)\delta^{(4)}(x-y)\right\}\\
-\epsilon^6\left\{\partial^6_x\left(c_6\delta^{(1)}(x-y)\right)+c_6 \delta^{(7)}(x-y)+(\partial_x c_6)\delta^{(6)}(x-y)\right\}\\
+\epsilon^6\left\{h \delta^{(3)}(x-y)+(\partial_x h) \delta^{(2)}(x-y)+\partial^2_x\left(h\delta^{(1)}(x-y)\right)\right\}\\
+\epsilon^6\left\{\partial^3_x\left((\partial^2_x g) \delta^{(2)}(x-y)\right)+\partial_x\left((\partial^3_x g)\delta^{(3)}(x-y)\right)+(\partial^2_x g)\delta^{(5)}(x-y)+(\partial^3_x g)\delta^{(4)}(x-y)\right\},
\end{split}
\end{equation}
where 
$c_2(u)$ is the central invariant and 
$c_4$ and $c_6$ are related to $c_2$ via the following equations:
\begin{equation}\label{c4asc2.eq}
c_4=-\frac{\d}{\d u}(c_2)^2,
\end{equation}
\begin{equation}\label{c6asc2.eq}
c_6=-\frac{1}{2}\frac{\d}{\d u}\left( c^2_2\; \frac{\d c_2}{\d u} \right),
\end{equation}
while $g$ is given by 
\begin{equation}\label{g.eq}
g=\frac{1}{2}\mathlarger{\int} \left\{\frac{3}{2}c_2^2\;  \frac{\d^3 c_2}{\d u^3}+\left(\frac{\d c_2}{\d u}\right)^3+\frac{19}{3}c_2 \; \frac{\d^2 c_2}{\d u^2}\;\frac{\d  c_2}{\d u} \right\}\; du
\end{equation}
and $h:=h_1+h_2+h_3+h_4$ and the $h_i$'s have the following expression in terms of the central invariant $c_2(u)$:

\begin{equation}
h_1=\,u_{xx}^{2}\left(\frac {97}{60}c_2 
\left(\frac{\partial^2 c_2}{\partial u^2}\right)^{2}+\frac{8}{3}\,\left(\frac{\partial c_2}{\partial u}\right)^{2}\frac{\partial^2 c_2}{\partial u^2} 
+\frac {21}{40}\, c_2^2\frac {\partial^4 c_2}{\partial u^4}+\frac {49}{15}\,c_2  
\left( \frac{\partial^3 c_2}{\partial u^3} \right) \frac{\partial c_2}{\partial u}\right)
\end{equation}
\begin{equation}
h_2=\,u_x^4 \left(
\begin{split}
\frac {254}{3}
\left(\frac{\partial c_2}{\partial u}\right) ^{2}\frac {\partial^4 c_2}{\partial u^4} +\frac {17}{5}\, \left(c_2\right)^{2}\frac {\partial^6 c_2}{\partial u^6} 
 +\frac{176}{3}\,c_2 \left(\frac {\partial^3 c_2}{\partial u^3}\right) ^{2}\\ +\frac {4018}{45}\, c_2 
\left( \frac {\partial ^4 c_2}{\partial u^4}\right) \frac{\partial^2 c_2}{\partial u^2}
+\frac {1684}{45}\,c_2 \frac{\partial^5 c_2}{\partial u^5} \frac{\partial c_2}{\partial u}+\frac {14512}{45}\,\left( \frac{\partial c_2}{\partial u}\right)  \left(  \frac{\partial^2 c_2}{\partial u^2}\right) \frac{\partial^3 c_2}{\partial u^3}
\end{split}
\right)
\end{equation}
\begin{equation}
h_3=u_{xxx}u_{x}\left( \frac{3}{10}\, c_2^2\frac{\partial^4 c_2}{\partial u^4} +\frac{2}{3}\, \left(\frac{\partial c_2 }{\partial u} \right) ^{2}\frac {\partial^2 c_2}{\partial u^2} +\frac{1}{15}\,c_2 \left( \frac{\partial^2 c_2}{\partial u^2}\right) ^{2}+\frac {28}{15}\,c_2 \left(\frac{\partial^3 c_2}{\partial u^3} \right) \frac{\partial c_2}{\partial u}\right)
\end{equation}
\begin{equation}
h_4=u_{xx}u_{x}^{2}\left(
\begin{split} \frac{139}{10}\, 
\left(\frac{\partial c_2}{\partial u} \right)  \left(  \frac{\partial^2 c_2}{\partial u^2} \right) ^{2}+\frac {178}{15}\,
\left(\frac{\partial c_2}{\partial u}\right) ^{2} \frac{\partial^3 c_2}{\partial u^3} 
+\frac {21}{20}\, c_2^2 \frac{\partial^5 c_2}{\partial u^5}\\
+\frac {259}{30}\,c_2  
\left(  \frac{\partial^4 c_2}{\partial u^4} \right)  \frac{\partial c_2}{\partial u} +13\,c_2 
\left( \frac{\partial^2 c_2}{\partial u^2} \right)  
\frac{\partial^3 c_2}{\partial u^3}
\end{split}
\right).
\end{equation}
\end{theorem}

Observe that the scalar pencil $\omega_{\lambda}$ is automatically homogeneous with Euler vector field $E=u\f{\partial}{\partial u}$. We want to check  that the equation \eqref{homPP} holds explicitly in this case for $k=1,2,3$ with arbitrary $D$ and with $d=0$. In particular, formula \eqref{homPP} gives in this specific case 
${\rm Lie}_E Q^{(2k)}_2=(kD-k-1)Q^{(2k)}_2$. On the other hand, since by formula \eqref{Lie} $${\rm Lie}_E Q^{(2k)}_{2}=\sum_{s\geq 0} (\partial^s_x u)\frac{\partial Q^{(2k)}_2}{\partial u_{(s)}}-2 Q^{(2k)}_2,$$
to check directly \eqref{homPP} is equivalent to show 
\begin{equation}\label{showhomPP}
\sum_{s\geq 0}(\partial^s_x u)\frac{\partial Q^{(2k)}_2}{\partial u_{(s)}}=(kD-k+1) Q^{(2k)}_2,
\end{equation}
 where $D$ is the degree of the central invariant $c_2$ as a function of $u$. 
 Now for $c_2:=u^D$ we have $Q^{(2)}_2=\partial^2_x\left( u^D\delta^{(1)}(x-y)\right)+u^D \delta^{(3)}(x-y)+(\partial_x u^D)\delta^{(2)}(x-y)$ and it is immediate to show that $\sum_{s\geq 0}(\partial^s_x u)\frac{\partial Q^{(2)}_2}{\partial u_{(s)}}=D Q^{(2)}_2$ using the identity \eqref{auxiliary2}. This confirms equation \eqref{showhomPP} in the case $k=1$. 
 
 For the case $k=2$, we have $Q^{(4)}_2=\partial^4_x \left(c_4 \delta^{(1)}(x-y)\right)+c_4\delta^{(5)}(x-y)+(\partial_x c_4)\delta^{(4)}(x-y)$. If $c_2=u^D$, then $c_4=-2D u^{2D-1}$ one gets $$Q^{(4)}_2=\partial^4_x \left(-2D u^{2D-1} \delta^{(1)}(x-y)\right)-2D u^{2D-1}\delta^{(5)}(x-y)+(\partial_x (-2D u^{2D-1}))\delta^{(4)}(x-y).$$
 Again applying \eqref{auxiliary2} we see that $\sum_{s\geq 0}(\partial^s_x u)\frac{\partial Q^{(4)}_2}{\partial u_{(s)}}=(2D-1)Q^{(4)}_2$, which is exactly \eqref{showhomPP} with $k=2$. 
 
 To check the homogeneity at $\epsilon^6$ is slightly more delicate. If $c_2=u^D$, then $c_6=-\frac{1}{2}D(3D-1)u^{3D-2}$, and consequently for the component of $Q^{(6)}_2$ given by $Q^{(6)}_{2,1}:=\partial^6_x\left(c_6\delta^{(1)}(x-y)\right)+c_6 \delta^{(7)}(x-y)+(\partial_x c_6)\delta^{(6)}(x-y)$ we have
 $$\sum_{s\geq 0}(\partial^s_x u)\frac{\partial Q^{(6)}_{2,1}}{\partial u_{(s)}}=(3D-2)Q^{(6)}_{2,1},$$
 which is \eqref{showhomPP} for $k=3$. Moreover, for the choice $c_2=u^D$, the function $g$ appearing in \eqref{g.eq} is equal to $a u^{3D-2}+b$, where $a$ is a suitable constant depending on $D$ while $b$ is a constant of integration. If we call $Q^{(6)}_{2,2}:=\partial^3_x\left((\partial^2_x g) \delta^{(2)}(x-y)\right)+\partial_x\left((\partial^3_x g)\delta^{(3)}(x-y)\right)+(\partial^2_x g)\delta^{(5)}(x-y)+(\partial^3_x g)\delta^{(4)}(x-y)$, again using \eqref{auxiliary2} we find
 $$ \sum_{s\geq 0}(\partial^s_x u)\frac{\partial Q^{(6)}_{2,2}}{\partial u_{(s)}}=(3D-2)Q^{(6)}_{2,2}.$$
Finally we need to check the homogeneity of $Q^{(6)}_{2,3}:=h \delta^{(3)}(x-y)+(\partial_x h) \delta^{(2)}(x-y)+\partial^2_x\left(h\delta^{(1)}(x-y)\right)$. Using the expressions for $h$ written above we find $h=\alpha u^2_{xx}u^{3D-4}+\beta u^4_x u^{3D-6}+\gamma u_{xxx}u_x u^{3D-4}+\rho u_{xx}u^2_{x} u^{3D-5}$, 
where $\alpha, \beta, \gamma, \rho$ are constants depending on $D$. This boils down to prove that 
$$ \sum_{s\geq 0}(\partial^s_x u)\frac{\partial h}{\partial u_{(s)}}=(3D-2) h,$$
which is immediate. Thus we have verified explicitly formula \eqref{homPP} up to $\epsilon^6$ for the non trivial bi-Hamiltonian deformations of the scalar pencil $\omega_{\lambda}$. 

\section{Conclusions}\label{conclusions}
In this  paper we have studied some general properties of
deformations of exact or homogeneous Poisson pencils
 of hydrodynamic type. In particular, 
we focused our attention on those characteristics of the fully deformed pencils that are inherited 
from properties of the dispersionelss limit (like exactness and homogeneity) coupled with suitable conditions about the central invariants.  

 In the case of exact Poisson pencil
we proved that their deformations can be reduced to a
suitable normal form
 via a Miura transformation. It turned out that each term 
of the deformation is annihilated by a sufficiently high
power of the
 operator ${\rm Lie}_e$. As a byproduct of this result we
showed that there exists a map between Poisson pencils
 with polynomial central invariants and Poisson pencil
with constant central invariants having the same
dispersionless limit.
 It would be interesting to extend this map to a wider
class of central invariants and to provide different maps performing the inverse task, namely starting from 
a pencil with constant central invariants and providing as an output pencils with polynomial central invariants 
(not constants). In principle, it is much more difficult to construct this last class of maps. 

Using similar ideas we also
constructed normal forms
 of deformations of homogeneous Poisson pencil of
hydrodynamic type having homogenous central invariants. In
this case it turned out that each term of the deformation is a homogeneous bivector of
specific degree. 

As  future investigations, it would be interesting to determine other characteristics of the fully deformed pencil
that are controlled by the dispersionless limit. In particular, one might ask if the requirement of the 
fulfillment of Virasoro constraints can be interpreted as a suitable property of dispersionless pencil.

{\bf Acknowledgments} The authors would like to thank the Department of Mathematics and Applications of the University of
Milano-Bicocca for the supportive atmosphere and Gregorio Falqui for stimulating discussions.  
In particular, A. A. would like to acknowledge the Department of Mathematics and Applications of the University of
Milano-Bicocca for the kind hospitality provided while this work was being written and the University of Toledo for 
support through selected funds.

\end{document}